\documentclass{amsart}

\usepackage{amsmath, amssymb, amsfonts, amscd, amsthm, mathrsfs}
\usepackage{tikz, tikz-cd}
\usepackage{comment}
\tikzset{%
    symbol/.style={%
        draw=none,
        every to/.append style={%
            edge node={node [sloped, allow upside down, auto=false]{$#1$}}}
    }
}
\usetikzlibrary{cd,arrows.meta,decorations.pathmorphing,shapes, positioning,calc, arrows, automata, matrix, positioning}
\usepackage{graphicx}
\usepackage{float}
\usepackage{epstopdf}
\usepackage[all]{xy}

\usepackage{url}
\makeatletter
\def\url@leostyle{%
  \@ifundefined{selectfont}{\def\UrlFont{\sf}}{\def\UrlFont{\small\ttfamily}}}
\makeatother
\newtheorem{theorem}{Theorem}[section]
\newtheorem{lemma}[theorem]{Lemma}
\newtheorem{proposition}[theorem]{Proposition}
\newtheorem{corollary}[theorem]{Corollary}

\newtheorem{example}[theorem]{Example}
\newtheorem{question}[theorem]{Question}

\newtheorem{definition}[theorem]{Definition}

\newtheorem{remark}[theorem]{Remark}

\newtheorem*{theorem*}{Theorem}
\newtheorem*{lemma*}{Lemma}
\newtheorem*{proposition*}{Proposition}
\newtheorem*{corollary*}{Corollary}
\newtheorem*{fact*}{Fact}
\newtheorem*{facts*}{Facts}
\newtheorem*{claim*}{Claim}

\newtheorem*{definition*}{Definition}
\newtheorem*{rmkdf*}{Remark and Definition}
\newtheorem*{remark*}{Remark}
\newtheorem*{example*}{Example}
\newtheorem*{examples*}{Examples}
\newtheorem*{exercise*}{Exercise}

\usepackage{graphicx} 

\newcommand{\N}{\mathbb{N}}
\newcommand{\Z}{\mathbb{Z}}
\newcommand{\Q}{\mathbb{Q}}

\newcommand{\pow}{\mathcal{P}}

\newcommand{\cM}{\mathcal{M}}
\newcommand{\cN}{\mathcal{N}}

\newcommand{\T}{\mathbb{T}}
\newcommand{\cat}[1]{\mathbf{#1}} 

\newcommand{\ev}{\mathrm{ev}}
\newcommand{\cA}{\mathcal{A}}
\newcommand{\cB}{\mathcal{B}}
\newcommand{\pres}{\mathbf{Pres}}

\usepackage[colorlinks=true, linkcolor=blue, citecolor=blue, urlcolor=blue]{hyperref}

\title{Rewriting Systems on Arbitrary Monoids}
\author{Eduardo Magalhães}
\date{August 2026}

\address{Department of Mathematics, University of Porto, Portugal}
\email{eduardomag79@gmail.com}

\begin{document}

\begin{abstract}
In this paper, we introduce monoidal rewriting systems (MRS), an abstraction of string rewriting in which reductions are defined over an arbitrary ambient monoid rather than a free monoid of words. This shift is partly motivated by logic: the class of free monoids is not first-order axiomatizable, so “working in the free setting” cannot be treated internally when applying first-order methods to rewriting presentations.

To analyze these systems categorically, we define $\cat{NCRS_2}$ as the 2-category of Noetherian Confluent MRS. We then prove the existence of a canonical 2-adjunction between $\cat{NCRS_2}$ and $\cat{Mon}$.

Finally, we study all MRS that present a given fixed monoid $M$. For this, we introduce Generalized Elementary Tietze Transformations (GETTs) and, among other results, we prove the property that any presentation of $M$ can be obtained from any other using GETTs is equivalent to $M$ being a free monoid, further strengthening the bridge between the algebraic behavior of monoids and rewriting theory.
\end{abstract}

\maketitle

\section{Introduction}
The study of rewriting systems has long occupied a central position at the intersection of theoretical computer science and algebra. Originating in the early 20th century with the work of Axel Thue \cite{power_thues_2013}, these systems were developed to address fundamental decidability questions such as the word problem for semigroups and monoids (and, indirectly, groups). Over the decades, the theory has matured into a robust framework for analysing algebraic structures by algorithmic means.

This field has been dominated by String Rewriting Systems (SRS), also known as semi-Thue systems. As detailed in the foundational work of Book and Otto "String rewriting systems" \cite{book_string-rewriting_1993}, an SRS is defined as a binary relation on a free monoid generated by some alphabet. When such a system is Noetherian and confluent, interpreting these relations as reduction rules allows one to rewrite complex words into unique irreducible forms, thereby offering a systematic approach to solving the word problem and other decidability problems.

Parallel to the development of SRS, the theory of Abstract Rewriting Systems (ARS) emerged to abstract certain key properties of reduction such as confluence, local confluence, termination and so on. The key idea behind this abstraction is that, although when working with SRS we are rewriting words on free monoids, the study of properties such as termination and confluence does not, in general, depend on the specific free monoidal structure present. As surveyed by Terese in "Term Rewriting Systems" \cite{terese_term_2003}, ARS theory provides general notions of reduction relations and normal forms that are present not only when working with string rewriting but also term rewriting and lambda calculus. However, despite this high level of abstraction and countless uses in computer science, applications to the study of monoids have traditionally remained grounded to the free context: reductions are almost exclusively defined as transformations between words in a free monoid $F_A$, for some alphabet $A$.

However, a structural limitation of classical rewriting systems is that they are intrinsically tied to free monoids. From a logical perspective, this creates a foundational mismatch:  Let $L=\{\cdot\}$ be the first-order language of monoids, and let $K=\{\cdot,R\}$ be its expansion by a designated binary relation symbol. Given a monoid, its presentations by rewriting can naturally be interpreted as structures in the richer language $K$, while many structural questions about the monoid itself are most naturally stated in $L$. The caveat is that, in the classical setting, rewriting is restricted to free monoids, which are not first-order axiomatizable. As such, when interpreting a $K$-structure as a rewriting system, “freeness” cannot be treated internally and must be imposed as an external semantic condition.

This observation suggests that, if one aims to use first-order/model-theoretic tools to study presentations of monoids, rewriting should be formulated intrinsically over arbitrary monoids (or at least over an elementary class of monoids).

Motivated by these considerations, we introduce Monoidal Rewriting Systems (from here onward abbreviated as MRS), a generalization of string rewriting systems in which reductions are defined directly over an arbitrary monoid, rather than over a free monoid of words. When string rewriting systems are used to present a monoid, the essential property that underpins the process is simply the ability to compose elements, that is, the presence of a monoidal structure. From this perspective, the syntactic representation provided by words in a free monoid is not strictly necessary. By defining rewriting relations intrinsically on a monoid, we decouple rewriting from its traditional syntactic setting and obtain a notion of reduction that is internal to the algebraic structure itself.

Furthermore, modern developments in category theory suggest that rewriting systems can be naturally understood not merely as static algebraic presentations, but as dynamic 2-dimensional structures, as exemplified by the theory of polygraphs and higher-dimensional word problems \cite{ara_polygraphs_2025,burroni_higher-dimensional_1993}. As emphasized in the work of Johnson and Yau \cite{johnson_2-dimensional_2021}, 2-categories provide a natural language to describe transformations between morphisms, making them an ideal setting for formalizing relationships between different rewriting systems. Motivated by this perspective, the present work not only abstracts the underlying algebraic setting of rewriting from free to arbitrary monoids, but also organizes these systems within a strict 2-categorical framework, so that generating systems and their equivalences can be studied using inherently higher-dimensional, categorical tools.

This paper is organized as follows:

Section 2 sets some notation and reviews some of the preliminaries necessary for the rest of this work. 

In section 3, we define a monoidal rewriting system (MRS) as a triple $(M,\cdot, R)$ where $(M,\cdot)$ is a monoid and $R \subseteq M \times M$ is a binary relation. We proceed by proving some small technical lemmas and results that are direct generalizations of basic known facts about SRS. One interesting fact about these lemmas is that their statements and proofs are essentially identical to those of their classical counterparts, which further strengthens the idea that the monoidal structure present in any SRS is the essential part that makes these lemmas work, and that the syntactic information that free monoids entail is not necessary at all. After this, we build the 2-category $\cat{NCRS_2}$ of Noetherian confluent MRS and explore the existence of a 2-adjunction between this 2-category and the 2-category of monoids $\cat{Mon}$. The proof of this 2-adjunction is quite cumbersome, and so we present it at the end in Appendix A.

Finally, in section 4, we answer the question: Given a fixed monoid $M$, what can be said about the family of MRSs that present $M$? For this, we introduce the notion of a generalized elementary Tietze transformation (GETT) and prove the property that we can get any presentation out of any other by performing a finite number of GETT is equivalent to the monoid $M$ being free.

\section{Preliminaries}\label{sec preliminaries}
We assume familiarity with the basics of Abstract Rewriting Systems (ARS) as detailed in \cite{terese_term_2003}. Given an ARS $(A, \to)$, we use $\to^*$ to denote the reflexive transitive closure of $\to$, $\leftrightarrow^*$ for the symmetric reflexive transitive closure of $\to$, and $\bar{u}$ to denote the unique normal form of an element $u$ in a Noetherian confluent ARS.

Recall that an ARS $(A, \to)$ is said to have the Church–Rosser property, if given $a,b \in A$ with $a \leftrightarrow^* b$, then there exists some $c \in A$ such that $a \to^*c$ and $b \to^* c$. The following is Theorem 2.1.5 in \cite{baader_term_1998}.

\begin{theorem*}
    Let $(A, \to)$ be a ARS. Then $(A, \to)$ is confluent if and only if it has the Church-Rosser property. 
\end{theorem*}

Given a set $A$, we use $F_A$ to denote the free monoid generated by $A$ with multiplication being given by concatenation.

\begin{definition*}
    A string rewriting system (SRS) is a pair $(A, R)$, where $A$ is a set and $R \subseteq F_A \times F_A$ is a binary relation.
\end{definition*}

For reasons that will become evident later on, we will denote a string rewriting system $(A,R)$ by $(F_A, R)$, where $R \subseteq F_A \times F_A$.

Given two words $u,v \in F_A$, we write $u \to_R v$ if there are some $a,b,x,y \in F_A$ such that $u = axb$, $v = ayb$ and $(x,y) \in R$.

If $(F_A, R)$ is a SRS, then $(F_A, \to_R)$ forms an ARS and so we say that a SRS $(F_A, R)$ is Noetherian [confluent] if $(F_A, \to_R)$ is Noetherian [confluent] as an ARS. Again, as $(F_A, \to_R)$ is an ARS, we define $\to_R^*$ and $\leftrightarrow_R^*$ to be the reflexive transitive closure and the symmetric reflexive transitive closure of $\to_R$ respectively.

Note that the relation $\leftrightarrow_R^*$ is a congruence on the monoid $F_A$. Given a monoid $M$ and a SRS $(F_A, R)$, we say that $(F_A, R)$ is a presentation of $M$ if $(F_A / \leftrightarrow_R^*) \cong M$. 

Recall the definition of elementary Tietze transformation (Definition 7.2.1 in \cite{book_string-rewriting_1993}):

\begin{definition*}
    Let $(F_A, R)$ be a string rewriting system. The following transformations of $(F_A, R)$ are called elementary Tietze transformations:
    \begin{enumerate}
        \item Given $u,v \in F_A$ with $u \leftrightarrow_R ^*v$, then $(F_A, R \cup \{(a,b)\})$ is obtained from $(F_A, R)$ by an elementary Tietze transformation of type 1;

        \item Given $(u,v) \in R$ with $u \leftrightarrow_{R\setminus \{(u,v)\}}^* v$, then $(F_A, R \setminus\{(u,v)\})$ is obtained from $(F_A, R)$ by an elementary Tietze transformation of type 2;

        \item Let $a \in F_A$ and $v$ be a new symbol, i.e. $v \not \in A$. Then $(F_{A \cup \{v\}}, R \cup \{(v,a)\})$ is obtained from $(F_A, R)$ by an elementary Tietze transformation of type 3;

        \item Let $a \in A$ and $u \in F_{A \setminus\{a\}}$ such that $(a,u) \in R$. Define a homomorphism $\phi_a: F_{A} \to F_{A \setminus\{a\}}$ by:
        $$\phi_a(b) = \begin{cases}
            b & \text{if $b \in F_{A \setminus\{a\}}$} \\
            u & \text{if $b = a$}
        \end{cases}$$
        Let $R_1 = \{(\phi_a(s), \phi_a(t)) : (s,t) \in R \setminus\{(a,u)\}\}$. Then $(F_{A \setminus\{a\}}, R_1)$ is obtained from $(F_A, R)$ by an elementary Tietze transformation of type 4.
    \end{enumerate}
\end{definition*}

Performing any type of elementary Tietze transformation does not change the monoid presented by a SRS (see Lemma 7.2.2 in \cite{book_string-rewriting_1993}). One of the most important uses of these transformations is the following theorem (see Theorem 7.2.4 in \cite{book_string-rewriting_1993}):

\begin{theorem*}[Tietze's Theorem]\label{tietze theorem}
    Let $(F_A,R)$ and $(F_B, L)$ be two finite presentations (i.e. the alphabets $A$ and $B$ are finite) of a monoid $M$. Then there exists a finite sequence of elementary Tietze transformations that transforms $(F_A, R)$ into $(F_B, L)$.
\end{theorem*}

Finally, we  briefly discuss the conventions about 2-categories adopted in this paper.

We work exclusively with strict 2-categories and strict 2-functors. We adopt the standard definition of strong transformations and 2-natural transformations as found in Definition 4.2.1 of \cite{johnson_2-dimensional_2021} . Specifically, a strong transformation $\alpha: F \to G$ consists of 1-cells $\alpha_A$ and invertible 2-cells $\alpha_f$ satisfying the naturality condition $G(f) \circ \alpha_A \cong \alpha_B \circ F(f)$. A strict 2-natural transformation is just a strong transformation where each component 2-cell $\alpha_f$ is the identity.

Finally, by 2-adjunction we mean the following:

\begin{definition*}
    Let $\cat C$ and $\cat D$ be strict 2-categories and consider two strict 2-functors $L: \cat{C} \to \cat{D}$ (left adjoint) and $R: \cat{D} \to \cat{C}$ (right adjoint);
        \[\begin{tikzcd}
\cat{C}\ar[r,bend left,"L",""{name=A, below}] & \cat{D} \ar[l,bend left,"R",""{name=B,above}] \ar[from=A, to=B, symbol=\dashv]
\end{tikzcd}\]

    A 2-adjunction is given by two strong transformations $\eta: 1_{\cat C} \to RL$ (unit) and $\varepsilon:LR \to 1_\cat{D}$ (counit) such that $\eta$ and $\varepsilon$ satisfy the triangle identities strictly:
    \begin{center}
        \begin{tikzcd}
                L \arrow[dr, "1_L", swap]\arrow[r, "L \eta"] &LRL \arrow[d, "\varepsilon L"]\\
                & L
            \end{tikzcd} \quad \quad \quad \begin{tikzcd}
                R \arrow[dr, "1_R", swap]\arrow[r, "\eta R"] &RLR \arrow[d, "R\varepsilon"]\\
                & R
            \end{tikzcd}
    \end{center}
\end{definition*}
\section{Monoidal Rewriting Systems}

We start by laying down the core definitions and proving some technical lemmas, which are simple generalizations of classical and fundamental results about string rewriting systems.

\begin{definition}
    A monoidal rewriting system (MRS) $\cM$ is a tuple $(M,\cdot, R)$ where $(M,\cdot)$ is a monoid and $R \subseteq M \times M$ is a binary relation.

    When the operation $\cdot$ on $M$ is implicit by context, we might write the MRS $(M,\cdot, R)$ as $(M,R)$.
\end{definition}

    \textbf{Note}: For the moment we postpone giving examples, in order to first introduce some further definitions.

    Given $a, b \in M$, we set $a \to_R b$ to mean that there exists $x,y,s,t \in M$ such that $a = xsy$, $b = xty$ and $(s,t) \in R$. This definition is analogous to the classical one we consider when working with SRS. The main difference is that now the elements are not words and multiplication is being taken internally to the monoid $M$.

    \begin{remark}\label{remark: arrow respects multiplication}
        The relation $\to_R$ is compatible with the monoid operation in the following sense: given $a,b,u,t \in M$, if $u \to _R t$ then $aub \to _R atb$.
    \end{remark}
    
    We use $\to_R^*$ to denote the reflexive and transitive closure of $\to _R$, meaning that $a \to_R^* b$ if and only if $a = b$ or if there exists $w_1,\ldots,w_n \in M$ with $n \geq 1$ such that $$a \to_R w_1 \to_R \ldots \to_R w_n \to_Rb$$

\begin{lemma}\label{lema tec 1}
    Let $\cM = (M,\cdot, R)$ be a MRS and $u,v,a,b\in M$. If $u \to _R^* a$ and $v \to _R^* b$ then $uv \to_R^* ab$.
\end{lemma}
\begin{proof}
    As stated in Remark \ref{remark: arrow respects multiplication}, the relation $\to_R$ is compatible with monoid multiplication. Because $u \to_R^* a$, we know that there exists $u_1,\ldots,u_n \in M$ such that $$u \to_R u_1 \to_R \ldots \to_R u_n \to_R a$$ As such, $$uv \to_R u_1v \to_R \ldots \to_R u_nv \to_R av$$

    So $uv \to_R^* av$. Repeating the same argument, we see that $av \to_R^* ab$, and therefore $uv \to_R^* ab$
\end{proof}

    We use $\leftrightarrow_R^*$ to denote the reflexive transitive symmetric closure of $\to _R$. An explicit description can be given as follows: Given $a,b \in M$, then $a \leftrightarrow_R^* b$ if and only if $a = b$ or if there are $w_1,\ldots,w_n \in M$, for some $n \geq 1$ such that $$w_i \to _R w_{i + 1} \text{ or } w_{i + 1} \to_R w_i$$ for $i = 0,\ldots,n$, where $w_0 := a$ and $w_{n+1} = b$.

    \begin{remark}\label{remark: definition derivation}
        If $a \leftrightarrow_R^* b$, then we will call the sequence $w_1,\ldots,w_n \in M$ described above a \emph{derivation} for $a \leftrightarrow_R^* b$.
    \end{remark}

\begin{lemma}\label{thue congruence}
    Let $(M,\cdot, R)$ be a MRS. Then equivalence relation $\leftrightarrow_R^*$ is a congruence.
\end{lemma}
\begin{proof}
    By definition, $\leftrightarrow_R^*$ is the reflexive, symmetric and transitive closure of the
one-step relation $\to_R$, hence it is an equivalence relation. It remains to show that it is
stable under multiplication on the left and on the right.

Now let $a,b \in M$ such that $a \leftrightarrow_R^* b$ and let  $x,y \in M$. If $a = b$, then it follows trivially that $xay \leftrightarrow_R^* xby$. On the other hand, let $w_1,\ldots,w_n \in M$ be a derivation for $a \leftrightarrow_R^* b$. Then $xw_1y,\ldots,xw_ny \in M$ is a derivation for $xay \leftrightarrow_R^* xby$
\end{proof}

\begin{definition}
    Let $N$ be a monoid. A presentation of $N$ is a pair $(\cA, \theta)$ where $\cA = (A, R)$ is a MRS that presents $N$ and $\theta: Q(\cA) \to N$ is a choice of monoid isomorphism, where $Q(\cA)$ denotes the quotient monoid $A/\leftrightarrow_R^*$. Furthermore, we denote by $q_\cA^\theta: A \twoheadrightarrow N$ the composition of the canonical projection $A \to Q(\cA)$ with the isomorphism $\theta: Q(\cA) \to N$.
\end{definition}

Note that although formally a presentation of a monoid $N$ is a pair $(\cA, \theta)$, we will sometimes just be interested in the MRS $\cA$ and disregard the choice of isomorphism $\theta$. If this is the case, we treat the presentation as just being given by the MRS $\cA$ whose quotient is isomorphic to $N$. Nevertheless, we shall make that isomorphism explicit when needed.

 When $(M, R)$ satisfies additional properties, which we will define now, there is a useful alternative characterization the quotient monoid $Q(M)$

\begin{definition}
    We say that the MRS $(A,R)$ is confluent, if the abstract rewriting system $(A,\to_R)$ is confluent. 

    We say that the MRS $(A,R)$ is Noetherian if every infinite chain $$w_1 \to_R w_2 \to_R w_3 \to\ldots$$ in $A$ eventually stabilizes.

    Given a MRS $(A, R)$, we say that an element $a \in A$ is irreducible if for any $v \in A$, $a \to_Rv$ implies that $a = v$.
\end{definition}

\textbf{Note:} When we are working with classical string rewriting system $(F_\Sigma, R)$, typically a word $u \in \Sigma ^*$ is said to be irreducible if there isn't a word $v \in \Sigma ^*$ such that $u \to_R v$. In particular, in this setting, if we can rewrite $u$ as $u$ (for example, if $(u,u) \in R$), then $u$ would not be considered irreducible. We move away from this notion of irreducibility because of the following: Let $M$ be any monoid with a $0$ element and consider any MRS $(M, R)$ with $R \neq \emptyset$. Given any $(a,b) \in R$, then $0 = 0\cdot a \cdot 0$ and $0 = 0 \cdot b \cdot 0$ so we always have $0 \to_R 0$, and as such $0$ would never be considered irreducible if we went with the classical definition. However, we might want $0$ to be irreducible in some scenarios, and as such, in this paper, an element is reducible only when we can rewrite it as a \emph{different} element. We also changed the definition of Noetherian for the same reason: If $M$ has a $0$ element, then any MRS $(M, R)$ with $R \neq \emptyset$ would never be Noetherian in the classical sense, as we can always form the infinite chain $0 \to_R 0 \to_R 0 \to \ldots$.

Even though we changed the definition of Noetherian and irreducible, the following familiar fact still holds.

\begin{lemma}\label{lemma: noetherian and confluent implies unique irreducible}
    Let $(M, R)$ be a Noetherian confluent MRS. Then, for every element $a \in M$, there exists a unique irreducible element $\bar a \in M$ such that $a \to_R^* \bar a$.
\end{lemma}
\begin{proof}
    Fix an element $a \in M$ and, for the sake of contradiction, assume that there is no irreducible element $v \in M$ such that $a \to _R^* v$. We inductively construct an infinite chain $w_1 \to_R w_2 \to_R w_3 \to\ldots$ as follows:

    \textbf{Base case:} We set $w_1$ to be $a$;

    \textbf{Inductive step:} For a given $n \geq 2$, assume that $w_{n-1}$ is already defined. By hypothesis, $w_{n-1}$ is not irreducible, since otherwise we would have $a \to_R^* w_{n-1}$ thus contradicting the fact that we can not rewrite $a$ as an irreducible element. As such, there must exist some $w_n$ such that $w_{n-1} \to_R w_n$ and $w_n \neq w_{n-1}$.

    These steps construct a sequence $w_1 \to_R w_2 \to_R w_3 \to\ldots$ that does not stabilize, thus contradicting the fact that $(M,R)$ is Noetherian. As such, for any $a \in M$, there exists some irreducible $v$ such that $a \to_R^* v$. Uniqueness follows easily from confluence.
\end{proof}

 Given a Noetherian confluent MRS $(M,R)$ and any $u \in M$, we denote by $\bar a$ the unique irreducible such that $a \to_R^* \bar a$, and call it the normal form of $a$. When multiple rewriting relations are present, to avoid ambiguity, we might denote $\overline{u}$ by $\overline{u}^R$.

Furthermore, given a MRS $(M, R)$, we denote the subset of all irreducible elements by $\overline M$. Again, when multiple rewriting relations are present, we might denote $\overline{M}$ by $\overline{M}^R$.

\begin{lemma}\label{lemma tecnico primeiro}
    Let $(M,R)$ be a Noetherian confluent MRS and $u,v\in M$. Then $\overline{u v} = \overline{\bar u  \bar v}$.
\end{lemma}
\begin{proof}
    By definition of normal form, we know that $u \to_R^* \bar u$ and $v \to_R^* \bar v$. Then $uv \to_R^* \bar u\bar v$ by Lemma \ref{lema tec 1}. Again, by the definition of normal form, we know that $\bar u \bar v \to_R^* \overline{\bar u \bar v}$. With these we can conclude that $$uv \to_R^* \bar u \bar v \to_R^* \overline{\bar u \bar v}$$
    Because $\overline{\bar u \bar v}$ is irreducible, by the uniqueness of the normal form, $\overline{u v} = \overline{\bar u  \bar v}$.
\end{proof}

We can give the set of all irreducible elements the structure of a monoid as follows.

\begin{lemma}
    Let $(M, R)$ be a confluent Noetherian MRS. Consider the binary operation on $\overline M$ given by: 
    \begin{align*}
        \bar \cdot: \overline M \times \overline M &\to \overline M   \\
        (u,v) &\mapsto \overline{u\cdot v}
    \end{align*}

    Then $(\overline M, \bar \cdot)$ is a monoid, which we call the monoid of irreducibles of $(M,R)$.
\end{lemma}
\begin{proof}
    We start by verifying that $\bar 1$ is the identity with respect to $\bar \cdot$. Let $a \in \overline M$ and note that, as $a$ is already irreducible, then $\bar a = a$. By Lemma \ref{lemma tecnico primeiro}, $$a \ \bar \cdot\  \bar 1 = \overline{a \cdot \bar 1} = \overline{\bar a \cdot \bar 1} = \overline{a \cdot 1} = \bar a = a$$
    The same argument shows that $\bar 1 \ \bar \cdot\  a = a$.

    As for the associativity, let $a,b,c \in \overline M$. Then:

    \begin{align*}
        a \ \bar \cdot\  ( b \ \bar \cdot\  c) &= a \ \bar \cdot\ (\overline{b\cdot c})
        =\overline{a \cdot(\overline{b\cdot c})}=\overline{ a \cdot(b\cdot c})\\
        &=\overline{ (a \cdot b)\cdot c}=\overline{ (\overline{a \cdot b})\cdot c}= (\overline{a \cdot b})\ \bar \cdot\  c =( a \ \bar \cdot\   b) \ \bar \cdot\  c \qedhere
    \end{align*}
\end{proof}

We now make precise the alternative description of $M / \leftrightarrow_R^*$ hinted at earlier.

\begin{lemma}
    Let $(M, R)$ be a Noetherian confluent MRS. Then the quotient monoid $M/\leftrightarrow_R^*$ is isomorphic to $(\overline{M}, \bar \cdot)$.
\end{lemma}
\begin{proof}
    Start by noting that, given $a \in M$, then $a \leftrightarrow_R^* \bar a$. Consider the map 
    \begin{align*}
        f:\overline{M} &\to M/\leftrightarrow_R^*\\ v&\mapsto [v]_R
    \end{align*}
    Where $[v]_R$ denotes the equivalence class of $v$ in $M/\leftrightarrow_R^*$.
    
    Let $a,b \in \overline{M}$. If $f(a) = f(b)$, then by definition of $f$, $a \leftrightarrow_R^* b$. By the Church–Rosser property, there exists some $c \in M$ such that $a \to_R^* c$ and $b \to_R^* c$. As both $a$ and $b$ are irreducibles, it follows that $a = c = b$.

    On the other hand, let $[a]_R$ be an equivalence class in $ M/\leftrightarrow_R^*$. Then $[a]_R \ = f (\bar a)$.

    Finally, let $a,b \in \overline{M}$ and note that 
    \begin{align*}
        f(a\ \bar \cdot \ b) &= [a\ \bar \cdot \ b]_R =  [\overline{a\cdot b}]_R\\
        &= [a\cdot b]_R= [a]_R \cdot [b]_R = f(a) f(b) \qedhere
    \end{align*}
\end{proof}

As stated earlier, we now explore some examples of presentations via MRSs. It is known that every monoid has a presentation given by a classical string rewriting system. As such, when we shift our focus towards presentations via MRSs, there are no `new' monoids that we can present this way that we couldn't previously. Nevertheless, the fact that we are able to rewrite direcly on a non-free monoid means that, for some monoids, we are able to give a very natural presentation via an MRS, whereas we would need a more convoluted string rewriting system to present it in a classical fashion.

\begin{example} \label{exemplo 1}
\ 

    \begin{enumerate}
   \normalfont \item Consider the MRS $(\N, +, R)$, where $R = \{(2,0)\}$. This MRS is Noetherian and confluent. The only two irreducible elements are $\overline{\N} = \{0,1\}$. For every element $n \in \N$, it's easy to see that its normal form will be $n \to_R^*( n \mod 2)$. As such, $(\N / \leftrightarrow_R^*) \simeq (\overline{\N}, \overline{+}) \simeq \Z/2\Z$;

    \item Let $M = \N \times \N$ with component-wise multiplication, and let $$R = \{\big((p,p), (1,1)\big) : \text{ $p$ is prime}\}.$$ Consider the MRS $(M, R)$ which is Noetherian and confluent. The irreducible elements are precisely the tuples $(a,b)$ where $\gcd(a,b) = 1$ and the monoid being presented by $(M,R)$ is $(\Q_{>0}, \times)$. Indeed, we may think of an irreducible element $(a,b)$ as denoting the fraction $a/b$.

    \item Let $(X,\tau)$ be a topological space and consider the monoid $(\pow X, \cup)$. Let $$R_0 = \{(U, \mathrm{cl}(U)): U \subseteq X\}$$
    where $\mathrm{cl}(\cdot)$ denotes the topological closure. Then, in $(\pow X, \cup , R_0)$, the irreducible elements are precisely the closed subsets of $X$. Furthermore, for every subset $U \subseteq X$ we have $U \to_{R_0} \mathrm{cl}(U)$, so $\pow X / \leftrightarrow_{R_0}^*$ is isomorphic to the monoid of closed subsets of $(X ,\tau)$ under union;

    \item Finally, we present an example from logic. Let $\Sigma$ be a single sorted first order language and let $\mathrm{ At}$ denote the set of atomic formulas.

    Consider a Horn theory $\T$ (see Section 1.2 in \cite{caramello_theories_2017} for the basic definitions and notation used here), and define the monoid $M = (\pow^{<\omega} \mathrm{At}, \cup)$ of finite subsets of atomic formulas. Note that a finite subset $\Delta \subseteq \mathrm{At}$ can be interpreted as the conjunction $\bigwedge_{\alpha \in \Delta}\alpha$.

    With the axioms of $\T$ in mind, we define a MRS $(M, \cup, R_\T)$ as follows:
    $$R_\T = \{(\Gamma_\phi,\Gamma_\phi \cup \Gamma_ \psi) : (\phi \vdash \psi) \in \T\}$$
    Where $\Gamma _\phi$ denotes the set of atomic formulas occurring in $\phi$ and $\Gamma_\psi$ is defined analogously.

    This definition means that given two finite subsets $\Delta, \Delta' \subseteq \mathrm{At}$, then $\Delta \to_{R_\T} \Delta'$ if and only if there exists some sequent $(\phi \vdash \psi) \in \T$ with $\Gamma_\phi \subseteq \Delta$ and $\Delta' = \Delta \cup \Gamma_\psi$.

    If the Horn theory $\T$ is finite, then $(M, \cup, R_\T)$ is Noetherian and confluent, and to the normal $\overline{\Delta}$ of a finite subset $\Delta \subseteq \mathrm{At}$, we call its Horn-closure. Then, in particular, an element $\Delta \in M$ is irreducible if and only if it is Horn-closed.
\end{enumerate}
\end{example}

So far we have defined MRS, proved some technical lemmas, and defined what the monoid of irreducibles associated with a Noetherian confluent MRS is. This way of assigning a monoid of irreducible elements to each Noetherian confluent MRS defines a map from the class of Noetherian confluent MRS to the class of monoids. What follows now is a way of building a suitable category where Noetherian confluent MRSs `live' such that this construction is actually functorial to the category of monoids. As we will see, there is also a way of assigning a canonical Noetherian confluent MRS to every monoid in a functorial way and, in fact, these two functors will form a (2-)adjunction.

We start by defining that a morphism between two MRSs is.

\begin{definition}
    Let $\cM = (M,R)$ and $\cN = (N,L)$ be two MRS. A MRS-homomorphism $\phi:\cM \to \cN$ is a monoid homomorphism $\phi:M \to N$ such that, for any $u,v \in M$, we have: $$(u,v) \in R \implies \phi(u) \to_L^* \phi(v)$$
\end{definition}

\textbf{Note:} Given a MRS-homomorphism $\phi:\cM \to \cN$, and $u,v \in M$, if $u \to_R^* v$ then  $\phi(u) \to_L^* \phi(v)$. Note also that the composition of two MRS-homomorphisms is still a MRS-homomorphism.

Before continuing we prove the following technical lemma.

\begin{lemma} \label{Lemma tecnico segundo}
    Let $\cM = (M, R)$ and $\cN = (N,L)$ be two Noetherian confluent MRS, $u\in M$ and $\phi:\cM \to \cN$ a MRS-homomorphism. Then: $\overline{\phi(u)}^L = \overline{\phi(\overline{ u}^R)}^L$.
\end{lemma}
\begin{proof}\ 
     By definition of normal form, we know that $u \to _R^* \overline{u}^R$. As $\phi$ is a MRS-homomorphism, then $\phi(u) \to_L^* \phi(\overline{u}^R)$. Now, consider the normal form of $\phi(\overline{u}^R)$: by definition $\phi(\overline{u}^R) \to _L^* \overline{\phi(\overline{u}^R)}^L$. With this we can conclude that $$\phi(u) \to_L^* \phi(\overline{u}^R) \to _L^* \overline{\phi(\overline{u}^R)}^L$$
    Because $\overline{\phi(\overline{u}^R)}^L$ is irreducible, by uniqueness of the normal form associated with any element, we have $\overline{\phi(u)}^L = \overline{\phi(\overline{u}^R)}^L$.
\end{proof}

Monoidal rewriting systems together with MRS-homomorphisms form a category, which we denote by $\cat{MRS}$. We use $\cat{NCRS}$ to denote the full subcategory of $\cat{MRS}$ of Noetherian confluent MRS.

We then give $\cat{NCRS}$ a (strict) $2$-categorical structure in the following way:

\begin{itemize}
    \item The $0$-cells are Noetherian confluent MRS;
    \item The $1$-cells are MRS-homomorphisms;
    \item Given two MRS-homomorphisms $f,g:(A, R) \to (B,L)$, there exists a unique 2-cell $\tau: f \Rightarrow g$ if and only if, for all $a \in A$, we have $f(a) \leftrightarrow_L^* g(a)$
\end{itemize}

In particular, given 1-cells $f$ and $g$, the category $\hom_{\cat{NCRS}} (f,g)$ is a thin groupoid since if there exists a 2-cell $\tau: \phi\Rightarrow\psi$, then there exists a 2-cell $\tau': \psi \Rightarrow \phi$ and uniqueness guarantees that they are inverses of each other.

Vertical composition of $2$-cells in this category works as follows:

Consider three parallel $1$-cells and two $2$-cells arranged in the following way

 \begin{center}
        \begin{tikzcd}
            (A,\cdot, R) \arrow[rr, "f", bend left=40, ""{name=lF, below}] \arrow[rr, "g" description, ""{name=lGup}, ""{name=lGdown, below}] \arrow[rr, "h", bend right=40, swap, ""{name=lHup}]& & (B,*,L) \arrow[Rightarrow, from=lF, to=lGup, "\ \alpha"] \arrow[Rightarrow, from=lGdown, to=lHup, "\ \beta",  end anchor={[yshift=3pt]}]
        \end{tikzcd}
    \end{center}

This means that for any $a \in A$, $f(a) \leftrightarrow_L^*g(a) \leftrightarrow_L^*h(a)$, so $\beta\circ \alpha:f \Rightarrow h$ is defined to be the unique 2-cell from $f$ to $h$.

As for horizontal composition, consider the following:

\begin{center}
        \begin{tikzcd}
            (A,\cdot, R) \arrow[rr, "f", bend left, ""{name=lF, below}]  \arrow[rr, "g", bend right, swap, ""{name=lHup}]& & (B,*,L) \arrow[Rightarrow, from=lF, to=lHup, " \alpha",  end anchor={[yshift=3pt]}] \arrow[rr, "f'", bend left, ""{name=lF', below}]  \arrow[rr, "g'", bend right, swap, ""{name=lH'up}] && (C,\times, W) \arrow[Rightarrow, from=lF', to=lH'up, " \alpha'",  end anchor={[yshift=3pt]}]
        \end{tikzcd}
    \end{center}

By the existence of the 2-cell $\alpha$, we know that for all $a\in A$, $f(a)\leftrightarrow_L^* g(a)$. As $f'$ is a MRS-homomorphisms, this implies that $f'f(a) \leftrightarrow_W^* f'g(a)$. By the existence of a 2-cell from $f'\Rightarrow g'$, we know that $f'g(a) \leftrightarrow_W^* g'g(a)$ so:

$$f'f(a) \leftrightarrow_W^* f'g(a) \leftrightarrow_W^* g'g(a)$$

So we define $\alpha'*\alpha$ to be the unique 2-cell from $f'f$ to $g'g$.

Then, the interchange law is rather straightforward to verify. Consider the following:

 \begin{center}
        \begin{tikzcd}
            (A,\cdot, R) \arrow[rr, "f", bend left=40, ""{name=lF, below}] \arrow[rr, "g" description, ""{name=lGup}, ""{name=lGdown, below}] \arrow[rr, "h", bend right=40, swap, ""{name=lHup}]& & (B,*,L) \arrow[Rightarrow, from=lF, to=lGup, "\ \alpha"] \arrow[Rightarrow, from=lGdown, to=lHup, "\ \beta",  end anchor={[yshift=3pt]}] \arrow[rr, "f'", bend left=40, ""{name=lF, below}] \arrow[rr, "g'" description, ""{name=lGup}, ""{name=lGdown, below}] \arrow[rr, "h'", bend right=40, swap, ""{name=lHup}]& & (C,\times,W) \arrow[Rightarrow, from=lF, to=lGup, "\ \alpha'"] \arrow[Rightarrow, from=lGdown, to=lHup, "\ \beta'",  end anchor={[yshift=3pt]}]
        \end{tikzcd}
    \end{center}

And consider $2$-cell $(\beta'*\beta) \circ (\alpha'*\alpha)$: $\beta'*\beta$ is the unique 2-cell from $g'g$ to $h'h$ while $\alpha'*\alpha$ is the unique 2-cell from $f'f$ to $g'g$, which means that $(\beta'*\beta) \circ (\alpha'*\alpha)$ is the unique 2-cell from $f'f$ to $h'h$.

On the other hand, consider $(\beta\circ \alpha)* (\beta'\circ \alpha')$: $\beta\circ \alpha$ is the unique 2-cell from $f$ to $h$, while $\beta'\circ \alpha'$ is the unique 2-cell from $f'$ to $h'$. As such, $(\beta\circ \alpha)* (\beta'\circ \alpha')$ is the unique 2-cell from $f'f$ to $h'h$. With this we can conclude that $$(\beta'*\beta) \circ (\alpha'*\alpha)= (\beta\circ \alpha)* (\beta'\circ \alpha')$$

When thinking of $\cat {NCRS}$ as a $2$-category, we will denote it by $\cat{NCRS_2}$. Additionally, we also consider $\cat{Mon}$ as a $2$-category where the categories $\hom_{\cat{Mon}}(f,g)$ are trivial for any two parallel morphisms $f$ and $g$, i.e. a locally discrete 2-category. In other words, there exists a unique 2-cell between two parallel monoid homomorphisms if and only if they are equal, and this unique 2-cell is the identity 2-cell.

The following lemma is the first step towards understanding how the category $\cat{NCRS_2}$ relates to $\cat{Mon}$.
\begin{lemma}
     Let $\cM = (M,\cdot, R)$ and $\cN = (N,*,L)$ be Noetherian confluent MRS, and $\phi:\cM \to \cN$ a MRS-homomorphism. Then, the map
     \begin{align*}
         \phi^\sharp : (\overline M, \bar \cdot) &\to (\overline N, \bar *) \\
         u &\mapsto \overline{\phi(u)}
     \end{align*}

     Is a monoid homomorphism. Additionally, given MRS-homomorphisms
     \begin{center}
         \begin{tikzcd}
             (M,\cdot,R) \arrow[r, "\phi"] & (N,*,L) \arrow[r, "\psi"] & (H,\times,W) 
         \end{tikzcd}
     \end{center}
     Then $(\psi \circ \phi)^\sharp = \psi^\sharp \circ \phi^\sharp$.
\end{lemma}
\begin{proof}
    Let $a,b \in \overline M$. Then, by definition $\phi^\sharp(a \ \bar \cdot\ b) = \overline{\phi(\overline{a\cdot b})}$. By Lemma \ref{Lemma tecnico segundo}, this is equal to $\overline{\phi(a \cdot b)}$ which is equal to $\overline{\phi(a) * \phi(b)}$. By Lemma \ref{lemma tecnico primeiro}, we have: $$\overline{\phi(a) * \phi(b)} = \overline{\overline{\phi(a)} * \overline{\phi(b)}} = \overline{\phi^\sharp(a) * \phi^\sharp(b)} = \phi^\sharp(a) \ \bar * \ \phi^\sharp(b)$$

    Additionally, let $\phi$ and $\psi$ be two composable MRS-homomorphisms. Given $u \in \overline M$, then
    \begin{align*}
        (\psi \circ \phi)^\sharp (u) &=\overline{(\psi \circ \phi) (u)}=\overline{\psi(\phi(u))}\\
        &=\psi^\sharp(\overline{\phi(u)})= \psi^\sharp(\phi^\sharp (u)) = (\psi^\sharp \circ \phi^\sharp)(u)\qedhere
    \end{align*}
\end{proof}
With this we can define a functor $$I: \cat{NCRS} \to \cat{Mon}$$
to the category of monoids, where given a confluent Noetherian MRS $(M,\cdot, R)$, then $I(M,\cdot, R)$ is its monoid of irreducibles, and given a MRS-homomorphism $\phi$, then $I\phi$ is the homomorphism $\phi^\sharp$ considered in the previous Lemma.

To extend $I$ to a strict 2-functor $\cat{NCRS_2} \to \cat{Mon}$, we first need to consider the following technical lemma.

\begin{lemma}
    Let $f,g:(A, R) \to (B,L)$ be two parallel morphism and $\tau: f \Rightarrow g$ a 2-cell between them. Then $If = Ig$.
\end{lemma}
\begin{proof}
    Recall that, by construction, the fact that there exists a 2-cell from $f$ to $g$ means that, for each $a \in A$, $f(a) \leftrightarrow_L^* g(a)$, and in particular, $\overline{f(a)}^L = \overline{g(a)}^L$. It follows therefore that $If = Ig$.
\end{proof}

\begin{remark}
    We can extend $I$ to a strict 2-functor $\cat{NCRS_2} \to \cat{Mon}$ by considering the functor $I:\hom_{\cat{NCRS}}(A,B) \to \hom_{\cat{Mon}}(\overline{A}, \overline{B})$ that sends the unique 2-cell $\tau:\phi  \Rightarrow\psi$ (if it exists) to the identity 2-cell $I\phi  \Rightarrow I\psi$.
    \end{remark}

\begin{example}
\ 

   \normalfont \begin{enumerate}
    \item Consider the Noetherian Confluent MRS $(\N, +, R)$ and $(\Z_2, +, \emptyset)$, where $R = \{(2,0)\}$. Consider the homomorphism \begin{align*}
        \phi: \N &\to \Z_2\\ n & \mapsto n \pmod 2
    \end{align*}

    Note that $\phi(2) \to_\emptyset \phi(0)$ as they are, in fact, equal. Thus $\phi$ is a MRS-homomorphism. Additionally, we have that $I\phi:I(\N,+,R) \to I(\Z_2, +, \emptyset)$ is an isomorphism.

    \item Using the notation fixed in Example \ref{exemplo 1} (5), consider two Horn theories $\T \subseteq \T'$, let $\iota:(M, \cup) \hookrightarrow (M, \cup)$ denote the identity map and note that this induces a MRS-homomorphism $$\iota: (M, \cup, R_\T) \hookrightarrow (M, \cup, R_{\T'})$$
    as $R_\T \subseteq R_{\T'}$.

    The functor $I$ then induces a monoid homomorphism $$I\iota: I(M, \cup, R_\T) \hookrightarrow I(M, \cup, R_{\T'})$$
    where, given a Horn-closed finite set of atomic formulas $\Delta$ with respect to the Horn theory $\T$, the map $I \iota$ will return the Horn-closure of $\Delta$ with respect to the theory $\T'$. Intuitively, this functor changes the Horn theory we are working with.
\end{enumerate}
\end{example}

Now, we define a functor $G: \cat{Mon} \to \cat{NCRS}$ that is, in a way, an `inverse' to $I$ (this will be made precise in Theorem \ref{biadju}). For now, consider the following definition:

\begin{definition}\label{def de G}
    Let $(M,\cdot)$ be a monoid and let $M^+$ denote the set $M \setminus\{1\}$. Let $\nu_M:M \hookrightarrow F_{M^+}$ be the injective function defined by $\nu_M(1) = \varepsilon$, where $\varepsilon$ denotes the empty word, and $\nu_M(m) = m$ where on the left hand side of the equality, $m$ is being considered as a letter on $F_{M^+}$.

    We define $G(M)$ as the MRS $(F_{M^+}, \oplus, R_M)$, where $\oplus$ denotes concatenation and $$R_M = \{(\nu_M(a) \oplus \nu_M(b), \nu_M(a\cdot b)): a,b \in M\}$$
\end{definition}

Before continuing I would like to make the following notes:

\textbf{Notes}: 
\begin{itemize}

    \item From now on, we will reserve the symbol $\oplus$ to always denote concatenation of words in free monoids. Additionally, given a monoid $M$, we will use $M^+$ to denote the set $M\setminus \{1\}$;

    \item In definition \ref{def de G}, the reason for first removing the element $1$ and only then taking the free monoid might seem artificial at first, but it is the most natural thing to do if we want $G$ to act as an `inverse' of $I$: more precisely, for any monoid $M$, we want to have the equality $IG(M) = M$. Imagine that we were to define $G$ as follows: 
    
    Let $G'(M)$ to be the MRS $(F_M, \oplus, R_M)$, and $R_M = \{(a\oplus b,c) : a\cdot b = c \text{ in $M$}\}$.
 
    And, for example, take $M$ to be $(\Z_2, +)$. Then $G'(\Z_2) = (F_M, \oplus, R_M)$, where $F_M$ is the free monoid in the alphabet $\{0,1\}$ and \begin{align*}
        R_M = \{&(00,0),(10, 1),\\&(01,1), (11,0)\}
    \end{align*}

    The irreducible elements in $G'(\Z_2)$ would then be $0$, $1$ and the empty word $\varepsilon$, meaning that $IG'(\Z_2)$ is a 3 element monoid with the identity element being $\varepsilon$, which in particular implies that $IG'(\Z_2) \not \simeq \Z_2$.

    As such, in order for $G$ to have the properties we desire, we start by removing the identity element from $M$ and we let $\varepsilon$ act as the identity element in $F_{M^+}$. When taking the monoid of irreducibles, $\varepsilon$ will play the same role in $IG(M)$ as $1$ plays in $M$.
\end{itemize}

\vspace{10pt}

\begin{remark}\normalfont \label{remark notation}
    To help simplify the notation and help with intuition when proving and stating the results, we will fix the following convention: We use the injective map $\nu_M$ identify $M =\nu_M(M)\subseteq F_{M^+}$.
Under this identification, elements of $M^+$ are viewed as letters
in $F_{M^+}$, while the identity element $1\in M$
is viewed as the empty word $\varepsilon$ .
For this reason, we will often omit $\nu_M$ from the notation and simply write $1$ for
$\varepsilon$ when no confusion can arise.

With this convention, we may write
\[
R_M=\{(a\oplus b,\,ab): a,b\in M\},
\]
understanding that the right-hand side $ab\in M$ is regarded as a word of
$F_{M^+}$ via the above identification (so, if $ab=1$, then $ab$ denotes
$\varepsilon$).
\end{remark}

\vspace{10pt}

Given two monoids $M$ and $N$ and an homomorphism $\phi:M \to N$, we can define a monoid homomorphism $G(\phi):F_{M^+} \to F_{N^+}$:

\begin{center}
    \begin{tikzcd}
        M^+ \arrow[d, hook, "\iota", swap]\arrow[r, "\phi|_{M^+}"] & N\arrow[d, "\nu_N",hook] \\
        F_{M^+}\arrow[r, dashed, "G\phi", swap] & F_{N^+}
        \end{tikzcd}
\end{center}

Where the map $G\phi$ is the unique homomorphism that exists by the universal property of free monoids applied to the map $\nu_N \circ \phi|_{M^+}:M^+ \to F_{N^+}$.

This homomorphism has the following explicit description: for $m_1,\ldots,m_n\in M^+$,
\[
G\phi(m_1 \oplus \cdots \oplus m_n)
= \nu_N(\phi(m_1)) \oplus \cdots \oplus \nu_N(\phi(m_n))
\]
Equivalently, using the identification $N=\nu_N(N)\subseteq F_{N^+}$ from above (see Remark \ref{remark notation}), one may write
\[
G\phi(m_1 \oplus \cdots \oplus m_n)=\phi(m_1)\oplus\cdots\oplus \phi(m_n)
\]
understanding that whenever $\phi(m_i)=1_N$, the term $\phi(m_i)$ contributes the empty word $\varepsilon$ (so it disappears under concatenation).

The following lemma states that this construction is functorial.

\begin{lemma}
    Let $\phi:M \to N$ be a monoid homomorphism. Then the map $G\phi: G(M) \to G(N)$ defined above is a MRS-homomorphism. Furthermore, for two composable monoid homomorphisms \begin{center}
         \begin{tikzcd}
             (M,\cdot) \arrow[r, "\phi"] & (N,*) \arrow[r, "\psi"] & (H,\times) 
         \end{tikzcd}
     \end{center} we have $G(\psi \circ \phi) = G(\psi) \circ G(\phi)$.
\end{lemma}
\begin{proof}
    The map $G\phi$ is already a monoid homomorphism $(F_{M^+},\oplus) \to (F_{N^+}, \oplus)$, so the only thing left to check in order to show that it is a MRS-homomorphism is that, given $u,v \in F_{M^+}$, then $$(u,v) \in R_M \implies G\phi(u) \to_{R_N}^* G\phi(v)$$

    Let $(u,v) \in R_M$. By definition, we know that $u$ is the string $a \oplus b$ for some $a,b \in M$ and $v = a \cdot b$ is a word of $F_{M^+}$ of length at most $1$ under the identification of Remark \ref{remark notation} (it is a letter if $v\neq 1_M$, and it is the empty word if $v=1_M$). As such, $$\phi(v) = \phi(a \cdot b) = \phi(a) * \phi(b)$$ meaning that $(\phi(a) \oplus \phi(b), \phi(v)) \in R_N$. Note, however, that $G\phi(u) =G\phi(a\oplus b) = \phi(a) \oplus \phi(b)$ and $G\phi(v) = \phi(v)$, so we have shown that $$ (G\phi(u), G\phi(v)) \in R_N$$
    which implies that  $G\phi(u) \to_{R_N}^* G\phi(v)$.

Now, let $\phi$ and $\psi$ be two composable homomorphisms. To show that $G(\psi \circ \phi) = G(\psi) \circ G(\phi)$, consider the diagram:

\begin{center}
        \begin{tikzcd}
        M^+ \arrow[d, hook, "\iota", swap]\arrow[r, "\phi"] \arrow[rr, bend left, "\psi \circ \phi"]& N\arrow[d, "\nu_N",hook] \arrow[r, "\psi"]& H \arrow[d, hook, "\nu_H"]\\
        F_{M^+}\arrow[r, dashed, "G\phi", swap] \arrow[rr, bend right, "G(\psi \circ \phi)", swap, dashed] & F_{N^+} \arrow[r, dashed, "G\psi", swap]& F_{H^*}
        \end{tikzcd}
\end{center}
Note that $\nu_H \circ (\psi \circ \phi) = (G\psi \circ G \phi) \circ \iota$. By uniqueness of $G(\psi \circ \phi)$ with respect to this property, we conclude that $G(\psi \circ \phi) = G(\psi) \circ G(\phi)$.
\end{proof}

As stated previously, from this lemma we can conclude that $G$ defines a functor $$G: \cat{Mon} \to \cat{NCRS}$$

Again, we also would like to consider $G$ as a (strict) 2-functor from $\cat{Mon}$ to $\cat{NCRS_2}$. This can be done by considering $$G:\hom_{\cat{Mon}}(M,N) \to \hom_{\cat{NCRS}}(G(M), G(N))$$ to be trivial on 2-cells, as $\hom_{\cat{Mon}}(M,N)$ is a discrete category.

\begin{theorem}\label{biadju}
    The pair $(G,I)$ forms a 2-adjunction, where $G$ is left adjoint and $I$ is right adjoint.
    \[\begin{tikzcd}
\cat{Mon}\ar[r,bend left,"G",""{name=A, below}] & \cat{NCRS_2} \ar[l,bend left,"I",""{name=B,above}] \ar[from=A, to=B, "\bot", phantom]
\end{tikzcd}\]

In particular, the unit $\eta:1_{\cat{Mon}} \to IG$ is the identity 2-natural transformation.
\end{theorem}
\begin{proof}
    The proof of this theorem is quite lengthy, although most of it consists of straightforward verifications that the unit and counit satisfy the necessary properties to yield a 2-adjunction. As to not disturb the flow of the text, you can find the proof at the end of the paper in Appendix A.
\end{proof}

\begin{example}\normalfont
Following again the notation of Example \ref{exemplo 1} (4), consider a Horn theory $\T$ and let $A = (M, \cup, R_\T)$. The monoid of irreducibles is then the monoid of Horn-closed finite sets of Horn-formulas with respect to $\T$, where multiplication is given by first taking the union and then taking the Horn-closure. Let the set of such Horn-closed finite sets if formulas be denoted by $X$.

    The MRS $GI(A)$ is given by $(F_X, \oplus, R_X)$, and the counit $$\varepsilon_A:GI(M, \cup, R_\T) \to (M, \cup, R_\T)$$

    Sends a string $\Delta_1 \oplus \ldots \oplus \Delta_n$ where each $\Delta_i$ is a finite subset of Horn-formulas that are Horn-closed, to their union $\Delta_1 \cup \ldots \cup \Delta_n$ in $M$.

    Here the Horn theory $\T$ and the rules $R_\T$ encode the logical syntax (one-step inferences with respect to $\T$), while the monoid $X$ is the corresponding algebraic semantics.  The functor $I$ passes from the syntactic MRS $A$ to the semantic monoid $X$, and $G$ reconstructs from $X$ the canonical convergent presentation $GI(A)$.
\end{example}

We conclude this section with a few remarks about Theorem \ref{biadju}.

First and foremost, the 2-categorical structures introduced here are necessary, in the sense that $G$ and $I$ do not form a classical adjunction between the categories $\cat{NCRS}$ and $\cat{Mon}$. Consider the following example.

\begin{example}\normalfont
    Consider the MRS $A = (\N, +, R)$ where $R = \{(2,0)\}$. Then there is no bijection between the sets $$\hom_\cat{NCRS}(G(\Z_2), A) \not\cong  \hom_\cat{Mon}(\Z_2, I(A))$$

    As $I(A) \cong \Z_2$, we have $|\hom_\cat{Mon}(\Z_2, I(A))| = 2$. 
    
    On the other hand, $G(\Z_2)$ is the MRS $(F_{\{1\}}, L)$ where $L = \{(11, \varepsilon),(1,1)\}$. As such, $|\hom_\cat{NCRS}(G(\Z_2), A)| = \aleph_0$, as any monoid homomorphism $F_{\{1\}} \to \N$ defines a MRS-homomorphism.
\end{example}

In general, the 2-adjunction $G \dashv I$ is best read as a universal property for the canonical presentation $G(M)$ of a monoid $M$, together with a precise description of how ``semantic" monoid maps lift to ``syntactic" maps of rewriting systems. Concretely, for every monoid $M$ and every Noetherian confluent MRS $A$, the 2-adjunction provides an equivalence of hom-categories
\[
\hom_{\cat{NCRS}_2}\bigl(G(M),A\bigr)\;\simeq\;\hom_{\cat{Mon}}\bigl(M,I(A)\bigr),
\]
natural in both variables (see \cite{lack_2-categories_2010}).

Because $\cat{Mon}$ is locally discrete, the right-hand side is a discrete category. In practice this rigidifies the left-hand side: the objects of $\hom_{\cat{NCRS}_2}(G(M),A)$ are classified, up to internal isomorphism, by monoid maps $M \to I(A)$, and any two 1-cells $G(M)\rightrightarrows A$ that induce the same monoid map are necessarily related by a unique 2-cell. Said differently, the 2-adjunction turns ``a monoid homomorphism out of $M$" into ``a map of rewriting systems out of the canonical presentation $G(M)$", and it does so uniquely up to the inevitable 2-cells in $\cat{NCRS}_2$.

An instance of this is the following: take $A = G(N)$ for another monoid $N$. Then the same equivalence reads
\[
\hom_{\cat{NCRS}_2}\bigl(G(M),G(N)\bigr)\;\simeq\;\hom_{\cat{Mon}}(M,N)
\]
As $IG(N) = N$. So on the image of $G$ the 2-category $\cat{NCRS}_2$ recovers $\cat{Mon}$ without loss of information: maps between canonical presentations are exactly monoid homomorphisms. 

\section{Generalized Elementary Tietze Transformations}

We now turn towards the following question: Given a fixed monoid $M$, what can be said about the MRSs that present it? What is the relationship between two presentations of the same monoid? 

To answer this question, we start by defining the main technical tool we will use, which is a generalization of elementary Tietze transformations, modified to be compatible with our generalized setting.

\begin{definition}
    Let $(A, R)$ be a MRS. The following procedures are called generalized elementary Tietze transformations (GETTs):

    \begin{enumerate}
        \item Let $X\subseteq A\times A$ such that, for every $(a,b) \in X$, we have $a \leftrightarrow_R ^*b$. Then $(A, R \cup X)$ is obtained from $(A, R)$ by a GETT of type 1 associated with the set $X$;

        \item Let $X \subseteq R$ such that, for every $(a,b) \in X$, we have $a \leftrightarrow_{R\setminus X}^* b$, then $(A, R \setminus X)$ is obtained from $(A, R)$ by a GETT of type 2 associated with the set $X$;

        \item Let $X$ be a set of new symbols, i.e. $X \cap A = \emptyset$ and let $f:X \to A$ be a function. Then $(A * F_X, R \cup \{(x,f(x)) : x \in X\})$ is obtained from $(A, R)$ by a GETT of type 3 associated with the set $X$ and the function $f$.

        \item Let $J \subseteq R$ be any subset, and let $(A_J, \cdot_J)$ denote the monoid $M/\leftrightarrow^*_J$ and $\pi_J: A\to A_J$ denote the canonical projection onto the quotient. Then $(A_J,\cdot_J, R_J)$, where $R_J = \{(\pi_J(a),\pi_J(b)) : (a,b)\in R\setminus J\}$ is said to be obtained from $(A,\cdot, R)$ by a type 4 GETT.
    \end{enumerate}

    Given two MRS $\cA$ and $\mathcal B$, we write $\cA \rightsquigarrow \mathcal B$ to mean that we can go from $\cA$ to $\mathcal B$ by performing a sequence of GETTs.
\end{definition}

\begin{remark}
    Starting with a Noetherian confluent MRS $(A,\cdot, R)$, if we pick a subset $J \subseteq R$ such that $(A,\cdot, J)$ is also confluent (note that it is always Noetherian no mater the choice of $J$), then there is an easy alternative description of the MRS obtained from this one by applying a type 4 GETT: 

    In this case, we can define $A_J$ to be the monoid of irreducibles $I(A,\cdot, J)$, and the set of rules $R_J$ is given by $R_J = \{(\overline{a}^J, \overline{b}^J) : (a,b)\in R\setminus J\}$
\end{remark}

Before moving on, we verify that performing a GETT of type 1,2, 3 or 4 does not change the monoid being presented by a given MRS.

\begin{lemma}\label{lemma: canonical isomorphism GETT}
    Let $\cA =(A, R)$ and $\mathcal B =(B, L)$ be two MRSs. If $\mathcal B$ is obtained from $\cA$ by performing a GETT of type 1,2,3 or 4, then there is a canonical monoid isomorphism $$Q(\cA) \overset{\sim}{\to} Q(\mathcal B)$$
\end{lemma}

   \begin{proof}
 We treat each GETTs separately. 

\smallskip
\noindent \textbf{(Type 1).}
If the GETT performed was of type 1, then the base set of each MRS is the same, i.e. $A = B$. Furthermore, every added rule as already derivable, meaning that $\leftrightarrow_R^*=\leftrightarrow^*_L$. Therefore we have an equality $Q(\cA) = Q(\mathcal B)$ and the canonical isomorphism is taken to be the identity.

\smallskip
\noindent \textbf{(Type 2).}
An argument analogous to that of a GETT of type 1 shows that, in this case, $Q(\cA) = Q(\mathcal B)$, so we take the canonical isomorphism to be the identity.

\smallskip
\noindent \textbf{(Type 3).} Fix a set $X$ with $X \cap A = \emptyset$ and a map $f:X \to A$. By definition $$\mathcal B = (A * F_X, R \cup \{(x,f(x)) : x \in X\}).$$
Let $\iota:A \hookrightarrow A * F_X$ denote the inclusion, and let $r:A*F_X \to A$ be the monoid homomorphism induced by setting: $$r|_A = \mathrm{id}_A, \quad r(x) = f(x) \text{ for } x \in X$$

Both this maps induce monoid homomorphisms \begin{center}
    \begin{tikzcd}
        Q(A) \arrow[r, "\hat\iota", bend left]& Q(A*F_X) \arrow[l, "\hat r", bend left]
    \end{tikzcd}
\end{center} which are mutually inverse of each other. The canonical isomorphism mentioned in the statement of the lemma is $\hat \iota$.

\smallskip
\noindent \textbf{(Type 4).} Let $J \subseteq R$ be any subset and consider the MRS $(A_J,\cdot_J, R_J)$ obtained from applying a type 4 GETT with respect to $J$.

Consider the map $\phi:A \to Q(A_J)$ defined as the composite of the canonical projections
\begin{center}
    \begin{tikzcd}
        A\arrow[r]& A/\leftrightarrow_J^* \arrow[r] & A_J/\leftrightarrow_{A_J}^*
    \end{tikzcd}
\end{center}

Note that $\leftrightarrow^*_{R_J}$ is a congruence on $A_J$, which itself is the quotient of $A$ by the congruence $\leftrightarrow_J^*$. Let $\pi_J: A\to A_J$ denote the canonical projection, and let $\tilde{X}$ denote the binary relation on $A$ defined as: $$(a,b)\in\tilde X \text{ if and only if } \pi_J(a) \leftrightarrow^*_{R_J} \pi_J(b) $$
By Theorem 6.20 in \cite{noauthor_course_nodate}, $\tilde X$ is a congruence on $A$ with $\leftrightarrow_J^* \subseteq \tilde X$. 

We start by showing: 

\textbf{Claim 1: $\leftrightarrow_R^* \subseteq \tilde X$}.

We prove this claim by dividing its proof into 3 intermediary claims. 

\textbf{Claim 1.1:} For any $a,b \in A$, if $a \to_R b$, then $(a,b)\in \tilde X$.

If this is the case, then there exists $x,y \in A$ and $(s,t)\in R$ such that $a = xsy$ and $b=xty$. If $(s,t)\in J$, then $a\leftrightarrow_J^* b$, from which it follows that $(a,b)\in \tilde X$, as we have already established that $\leftrightarrow_J^*\subseteq \tilde X$. On the other hand, if $(s,t)\in R\setminus J$, then by definition, $(\pi_J(s), \pi_J(t)) \in R_J$, from which it follows that $\pi_J(a)\to_{R_J} \pi_J(b)$. From here we conclude that $\pi_J(a)\leftrightarrow^*_{R_J} \pi_J(b)$ which, by the definition of $\tilde X$, implies that $(a,b) \in \tilde X$.

\textbf{Claim 1.2:} For any $a,b \in A$, if $a \to_R^* b$, then $(a,b)\in \tilde X$.

This follows trivially from the transitivity of $\tilde X$ together with Claim 1.1.

\textbf{Claim 1.3:}  For any $a,b \in A$, if $a \leftrightarrow_R^* b$, then $(a,b)\in \tilde X$.

Let $a = w_0, \ldots, w_n = b$ be a derivation of $a \leftrightarrow_R^* b$ (see Remark \ref{remark: definition derivation}). By claim 1.2, it follows that for each $i < n$, then $(w_i, w_{i + 1}) \in \tilde X$ or $(w_{i + 1}, w_i) \in \tilde X$. The fact that $\tilde X$ is symmetric implies that  $(w_i, w_{i + 1}) \in \tilde X$, for each $i < n$, and by transitivity if follows thus that $(a,b)\in \tilde X$.

\vspace{10pt}

As such, if $a\leftrightarrow_R^* b$ in $A$, we have that $(a,b) \in \tilde X$, which by definition of $\tilde X$, means that $\pi_J(a) \leftrightarrow^*_{R_J} \pi_J(b)$. In particular, from this we conclude that if  $a\leftrightarrow_R^* b$ then $\phi(a) = \phi(b)$. In particular, this proves that the following map is a surjective monoid homomorphism \begin{align*}
    \Phi: Q(A) & \to Q(A_J) \\
    [a]_R & \mapsto [[a]_J]_{R_J}
\end{align*}

This will be the canonical isomorphism mentioned in the statement of the theorem.

To show that this is injective, we start with the following claim.

\textbf{Claim 2:} $\tilde X \subseteq \leftrightarrow_R^*$

Let $(a,b) \in \tilde X$. Note that, by definition of $\tilde X$, it suffices to show that $\pi_J(a) \leftrightarrow^*_{R_J} \pi_J(b)$ implies $a \leftrightarrow^*_R b$. 

In order to prove this, we divide the proof into 3 intermediary claims as well.

\textbf{Claim 2.1:} if $\pi_J(a)\to_{R_J} \pi_J(b)$, then $a \leftrightarrow^*_R b$.

The fact that $\pi_J(a)\to_{R_J} \pi_J(b)$ means that there exists $x,y \in A$ and $(s,t) \in R\setminus J$ such that $$\pi_J(a) = \pi_J(x)\pi_J(s)\pi_J(y) = \pi_J(xsy)$$
$$\pi_J(b) = \pi_J(x)\pi_J(t)\pi_J(y) = \pi_J(xty)$$

From the equality $\pi_J(a) = \pi_J(xsy)$, we conclude that $a \leftrightarrow_J^* xsy$. Analogously, we also have that $b \leftrightarrow_J^* xty$. Because $(s,t) \in R\setminus J$, we conclude that $xsy\leftrightarrow ^*_R xty$. From this, it follows that $a\leftrightarrow_R^* b$.

\textbf{Claim 2.2:} if $\pi_J(a)\to^*_{R_J} \pi_J(b)$, then $a \leftrightarrow^*_R b$.

This follows trivially from the transitivity of $\leftrightarrow_R^*$ together with Claim 2.1.

\textbf{Claim 2.3:} if $\pi_J(a)\leftrightarrow^*_{R_J} \pi_J(b)$, then $a \leftrightarrow^*_R b$.

Let $a = w_0, \ldots, w_n = b$ be a derivation for $\pi_J(a) \leftrightarrow_{R_J}^* \pi_J(b)$. By claim 2.2, it follows that for each $i < n$, then $w_i\leftrightarrow_R^* w_{i + 1}$ or $w_{i + 1} \leftrightarrow_R^* w_i$, from which if follows that $a \leftrightarrow_R^*b$.

\vspace{10pt}

To conclude the proof, let $a,b\in A$ such that $[[a]_J]_{R_J} = [[b]_J]_{R_J}$. We can rewrite this last equality as $\pi_J(a) \leftrightarrow^*_{R_J} \pi_J(b)$, i.e. $(a, b) \in \tilde X$. From the last claim, follows that $a \leftrightarrow_R^* b$, from which we conclude that $[a]_R = [b]_R$.
\end{proof}

By chaining the canonical isomorphisms described in this lemma, we obtain the following corollary.

\begin{corollary}\label{corollary: square commutes}
    Fix two MRSs $\cA = (A, R)$ and $\mathcal B = (B,L)$. Let $\sigma: \cA \rightsquigarrow \mathcal B$ be a sequence of GETTs. Then $\sigma$ determines a monoid homomorphism $c_\sigma :A \to B$ and an induced isomorphism $\bar c_\sigma: Q(\mathcal A) \to Q(\mathcal B)$ such that the following square commutes
    \begin{center}
        \begin{tikzcd}
             A \arrow[r, "c_\sigma"]\arrow[d, "q_A", swap] & B \arrow[d, "q_B"]\\ Q(A)\arrow[r, "\bar c_\sigma", swap] & Q(B)
        \end{tikzcd}
    \end{center}
\end{corollary}
\begin{proof}
    Let $\sigma: \cA \rightsquigarrow \mathcal B$ denote a sequence of GETTs:
    $$\cA = \mathcal D_0  \Rightarrow \ldots \Rightarrow \mathcal D_n = \mathcal B$$ where each double arrow represents a specific GETT of type 1,2,3 or 4 and each $\mathcal D_{i+1}$ is a MRS obtained from $\mathcal D_{i}$ by performing that respective GETT.
    
    For each type of GETT in this sequence, define a monoid homomorphism $c_i:D_i \to D_{i+1}$ by $$c_i =\begin{cases}
        \mathrm{id_{D_i}}, & \text{for Types 1 and 2}\\
        D_i \hookrightarrow D_i *F_X, & \text{for Type 3}\\
        \pi_J: D_i\twoheadrightarrow (D_i)_J, & \text{for Type 4}
    \end{cases}$$

    We define $c_\sigma:A \to B$ as being the composite of all these homomorphism for each GETT in $\sigma$. Analogously, we define $\bar c_\sigma$ to be the composite of the canonical isomorphisms $$Q(\cA) = Q(\mathcal D_0)  \to \ldots \to Q(\mathcal D_n) = Q(\mathcal B)$$ as defined in Lemma \ref{lemma: canonical isomorphism GETT}. Consider the following diagram:
    \begin{center}
        \begin{tikzcd}
             D_0 \arrow[r, "c_0"]\arrow[d, "q_{Q_0}", swap] & D_1 \arrow[d, "q_{D_1}"]\ \arrow[r, "c_1"] &\ldots & \ldots \arrow[r, "c_{n-1}"] & D_n \arrow[d, "q_{D_n}"]\\ 
             Q(D_0)\arrow[r, "\bar c_0", swap] & Q(D_1)  \arrow[r, "\bar c_1", swap] &\ldots&\ldots\arrow[r, "\bar c_{n-1}", swap] & Q(D_n)
        \end{tikzcd}
    \end{center}
    Each square in this diagram is commutative as per the constructions of the canonical isomorphisms in Lemma \ref{lemma: canonical isomorphism GETT}, and therefore the most outer square also commutes, which concludes the proof.
\end{proof}

The next theorem gives translates the rewriting property of $\cB$ being obtainable from $\cA$ via a sequence of GETT into an algebraic property of the monoids involved in the rewriting systems $\cA$ and $\cB$.

\begin{theorem}\label{theorem: if square commutes then GETT sequence}
    Fix two MRSs $\cA = (A, R)$ and $\mathcal B = (B,L)$. The following are equivalent:
    \begin{enumerate}
        \item There exists a sequence of GETTs $\cA \rightsquigarrow \mathcal B$;

        \item There exists a monoid homomorphism $f:A \to B$, and an isomorphism $\alpha: Q(\cA) \to Q(\cB)$ such that the following square commutes: 
        \begin{center}
        \begin{tikzcd}
             A \arrow[r, "f"]\arrow[d, "q_A", swap] & B \arrow[d, "q_B"]\\ Q(A)\arrow[r, "\alpha", swap] & Q(B)
        \end{tikzcd}
    \end{center}
    \end{enumerate}
\end{theorem}
\begin{proof}
    We already proved that (1) implies (2) in Corollary \ref{corollary: square commutes}. 

    On the other hand, assume that (2) is true. Without loss of generality, assume that $B\cap A = \emptyset$. Define $d:B^+ \to A$ to be a function such that, for each $b \in B^+$, the following equality holds:
    \begin{align}\label{property d}
        \alpha q_A d(b) = q_B(b).
    \end{align}

    Such a function exists by the surjectivity of $\alpha q_A$.

    Let $W'$ be the set $\{(\underline{b},d(b)) : b \in B^+\}$, where $\underline{b}$ denotes the element $b$ though of as a letter in $F_{B^+}$, and consider the new MRS $$(A* F_{B^+}, R \cup W')$$ obtained from applying a type 3 GETT.

    The next step is to try to add the rules $R_B$ to our MRS using a type 1 GETT. For that, we must first prove the following claim.

    \textbf{Claim 1:} For each $b_1,b_2 \in B^+$, $\underline{b_1}\ \underline{b_2} \leftrightarrow^*_{R\cup W'} \underline{b_1b_2}$, where the underline denotes an element seen as a letter in $F_{B^+}$.

    Start by noting that $$\underline{b_1} \ \underline{b_2} \to^*_{W'} d(b_1) d(b_2)$$  $$\underline{b_1b_2} \to_{W'} d(b_1b_2)$$

    As such, it suffices to show that $d(b_1)d(b_2)\leftrightarrow^*_{R\cup {W'}} d(b_1b_2)$

    For that, consider the following equalities: $$\alpha q_A d(b_1b_2) = q_B(b_1b_2) = q_B(b_1) q_B(b_2) = \alpha q_A d (b_1) \cdot \alpha q_A d (b_2) = \alpha q_A (d(b_1) d(b_2))$$

    By bijectivity of $\alpha$, it follows that $q_A d(b_1b_2) = q_A (d(b_1) d(b_2))$. From this we can conclude that $d(b_1b_2)\leftrightarrow_R^* d(b_1)d(b_2)$.

    \vspace{10pt}
    As per the claim we just proved, we can then apply a type 1 GETT and obtain: $$(A* F_{B^+}, R \cup {W'}\cup R_B)$$
    
    The next step is to add the rules in $L$ using type 1 GETTs. Again, for that, we start by proving the following claim.

    \textbf{Claim 2}: For each $(a,b) \in L$, we have that $\underline{a} \leftrightarrow^*_{R\cup {W'} \cup R_B} \underline b$ 

    Start by noting that, if $(a,b) \in L$, then in particular $a \leftrightarrow_L^* b$, and therefore $q_B(a) = q_B(b)$. By \eqref{property d}, we obtained therefore the following chain of equivalences
    \begin{align*}
        q_B(a) = q_B(b) &\iff \alpha q_Ad(a) = \alpha q_A d(b)\\
        &\iff q_A d(a) = q_Ad(b) \\
        &\iff d(a) \leftrightarrow_R^* d(b)
    \end{align*}

    Additionally, note that $\underline{a} \to_{W'} d(a)$ and $\underline{b} \to_{W'} d(b)$, from which it follows that $\underline{a} \leftrightarrow^*_{R\cup {W'} \cup R_B} \underline b$

    \vspace{10pt}

    This claim grantees that we may add $L' = \{(\underline{a}, \underline{b}) : (a,b) \in L\}$ to the MRS via a type 1 GETT, thus obtaining: $$(A* F_{B^+}, R \cup {W'}\cup R_B \cup L').$$

    In this next step of the proof, we use a type 4 GETT to applying every rule in $R_B$. The resulting MRS is $$(A* B, R \cup W\cup L),$$ where now $W = \{(b,d(b)) : b \in B^+\}$.

    \textbf{Claim 3:} For each $a \in A$, we have $a \leftrightarrow^*_{R \cup W\cup L} f(a)$.

    Start by considering, for any $a\in A$, the element $df(a)$. Using \eqref{property d}, we obtain the following equality
    $$\alpha q_A df(a) = q_Bf(a).$$

    From the fact that $q_B f =\alpha q_A$, we can conclude the following $$\alpha q_A df(a) = \alpha q_A(a)$$ which, from the bijectivity of $\alpha$, gets us that $q_A df(a) = q_A(a)$ and therefore $a \leftrightarrow_R^* df(a)$. Using the rewriting rules in $W$, we also have that $df(a) \leftrightarrow_W^* f(a)$, from which the claim follows.

    \vspace{10pt}

    With this claim, we can add the set of rules $Z = \{(a,f(a)) : a \in A\}$ to the MRS using a type 1 GETT, thus obtaining $$(A* B, R \cup W\cup L \cup Z).$$ 

    The next step in the proof is to remove the rules in $W$ using a type 2 GETT. For that, the following claim is needed.

    \textbf{Claim 4:} For each $b \in B^+$, we have $b \leftrightarrow^*_{R\cup L \cup Z} d(b)$.

    Start by noting that $d(b) \to_Z fd(b)$. Now by \eqref{property d}, we know that $$\alpha q_A d(b) = q_B(b).$$ Using the fact that $\alpha q_A = q_B f$ on the left hand side of this equality, we conclude that $$q_B f d(b) = q_B (b)$$ from which it follows that $fd(b) \leftrightarrow_L^* b$, thus proving the claim.

    \vspace{10pt}

    As such, using a type 2 GETT to remove the set $W$, we obtain the MRS: $$(A* B,  R \cup L \cup Z).$$ 

    Now we wish to remove the set of rules $R$.

    \textbf{Claim 5:} For each $(a,b) \in R$, we have $a\leftrightarrow_{L \cup Z} b$.

    Note that, $a \leftrightarrow_R b$, which in particular means that $q_A(a) = q_A (b)$. From here we obtain the following chain of equivalences:
    \begin{align*}
        q_A(a) = q_A(b) &\iff \alpha q_A(a) = \alpha q_A (b)\\
        &\iff q_B f(a) = q_Bf(b) \\
        &\iff f(a) \leftrightarrow_L^* f(b)
    \end{align*}
    Finally, from the fact that $a \to_Z f(a)$ and $b \to_Z f(b)$, the claim follows.

    \vspace{10pt}

    Apply a type 2 GETT to remove the set $R$, thus obtaining the MRS:  $$(A* B,L \cup Z).$$ 

    The final step of the proof is to apply a type 4 GETT to the set of rules in $Z$. After doing so, we are left with the MRS $(B, L)$, which concludes the proof of the theorem.
\end{proof}

Theorem \ref{theorem: if square commutes then GETT sequence} is the main technical tool we will use to derive the results that follow. We start with the following corollary.

\begin{corollary}\label{corollary: G(M) is weakly initial}
    Let $M$ be a monoid and $\cA$ be a presentation of $M$. Then $G(M) \rightsquigarrow \cA$.
\end{corollary}
\begin{proof}
    Choose an isomorphism $\theta: Q(A)\to M$ and consider the following diagram
    \begin{center}
        \begin{tikzcd}
            G(M)\arrow[d, "q", swap] & A \arrow[d, "q_A"] \\
            M\arrow[r, "\theta^{-1}", swap] & Q(A)
        \end{tikzcd}
    \end{center}
    Let $\hat f:M \to A$ be a function such that $$q_A f(m) = \theta^{-1} q(\underline m)$$ where $\underline m$ denotes the element $m$ thought of as a letter in $F_{M^+}$. Such a function exists due to the surjectivity of $q_A$. Additionally, such a function extends to a monoid homomorphism $f:F_{M^+} \to A$ which satisfies $q_A f = \theta^{-1} q$ by construction. By Theorem \ref{theorem: if square commutes then GETT sequence}, we conclude that $G(M)\rightsquigarrow \cA$.
\end{proof}

\begin{remark}
    We can view this result categorically in the following way: for a given monoid $M$, let $\pres(M)$ be the category whose objects are presentations of $M$, and whose arrows are finite sequences of GETTs, with composition being given by juxtaposition of such sequences (we consider the empty sequences as being identities for each presentation).

    Then, Corollary \ref{corollary: G(M) is weakly initial} is precisely stating that $G(M)$ is weakly initial in $\pres(M)$.

    On the other hand, given a monoid $M$, let $D(M)$ denote the MRS $(M,\cdot, \emptyset)$. Then $D(M)$ is (trivially) a presentation of $M$ and, in particular, starting with a MRS $\cA$ that presents $M$, we have that $\cA \rightsquigarrow D(M)$ by performing a single GETT of type 4. 

    As such, given a monoid $M$, $D(M)$ is weakly terminal in $\pres(M)$.
\end{remark}

This property if $G(M)$ being weakly initial in $\pres(M)$ is not a property that emerges because of the way we constructed $G(M)$, but simply because $G(M)$ is a classical string rewriting system, as we will see now.

\begin{proposition}\label{proposition: SRS are initial}
    Let $M$ be a monoid and $(F_A, R)$ a string rewriting system that presents it, for some alphabet $A$. Then $(F_A, R)$ is weakly initial in $\pres(M)$.
\end{proposition}
\begin{proof}
    Let $q: F_A \to Q(F_A)$ denote the canonical projection, fix an isomorphism $\theta: Q(A) \to M$ , let $\nu_M:M \hookrightarrow F_{M^+}$ be the canonical inclusion (recall Remark \ref{remark notation}) and $q_M: Q(M) \to M$ the canonical projection.

    Let $\hat f:A \to F_{M^+}$ denote the composition
    \begin{center}
        \begin{tikzcd}
            A \arrow[r, "q_A"]& Q(A) \arrow[r, "\theta"]& M \arrow[r, "\nu_M"] & F_{M^+}
        \end{tikzcd}
    \end{center}
    By the universal property of free monoids, this extends to a monoid homomorphism $f:F_A \to F_{M^+}$.

    Let $\underline{a} \in F_A$ denote the element $a\in A$ thought of as a letter in $F_A$ and note that $$qf(\underline{a}) = \underbrace{q_M\nu_M}_{1_M}\theta q_A(\underline{a}) = \theta q_A(\underline{a})$$
    From this fact, it follows that the following square commutes
    \begin{center}
        \begin{tikzcd}
            F_A \arrow[r, "f"]\arrow[d, "q_A", swap]& G(M)\arrow[d, "q_M"] \\ Q(F_A)\arrow[r, "\theta", swap] & M
        \end{tikzcd}
    \end{center}
    By Theorem \ref{theorem: if square commutes then GETT sequence}, $F_A \rightsquigarrow G(M)$, and from the fact that $G(M)$ is weakly initial in $\pres(M)$, it follows that so is $(F_A,R)$.
\end{proof}

Note that one of the key aspects of the proof of Proposition \ref{proposition: SRS are initial} is that there exists a function $\nu_M : M \to F_{M^+}$ that is a section for $q_M$, in the sene that $q_M \nu_M = 1_M$. This motivates the following definition.

\begin{definition}
    Let $M$ be a monoid. A presentation $(\cA, \theta)$ is said to be split if $q_\cA^\theta : A \twoheadrightarrow M$ admits a monoid section, in the sense that there exists a monoid homomorphism $s:M \to A$ such that $q_\cA^\theta s = 1_M$.
\end{definition}

The first observation is that being split is independent of the choice of isomorphism made.

\begin{lemma}
    Let $(\cA,\alpha)$ and $(\cA,\beta)$ be two presentations of the same monoid such that $(\cA,\alpha)$ is split. Then so is $(\cA,\beta)$.
\end{lemma}
\begin{proof}
    Let $s_\alpha: M \to A$ be a monoid homomorphism such that $\alpha q_As_\alpha = 1_M$ (and note that, in particular, $q_A s_\alpha = \alpha^{-1}$). Define $s_\beta := s_\alpha  \alpha \beta^{-1}$.
    \begin{center}
        \begin{tikzcd}
            A \arrow[d, "q_A" description]\\ Q(A) \arrow[d, bend left, "\beta" description] \arrow[d, bend right, "\alpha" description] \\ M \arrow[uu, bend left = 60, "s_\alpha"] \arrow[uu, bend right= 60, "s_\beta", swap]
        \end{tikzcd}
    \end{center}

    By construction of $s_\beta$, we have:
    \begin{align*}
        \beta q_A s_\beta =\beta (q_A s_\alpha ) \alpha \beta^{-1} = \beta\alpha^{-1}\alpha\beta^{-1} = 1_M
    \end{align*}
    which concludes the proof.
\end{proof}

\begin{remark}
    This lemma motivates the following convention: Given a MRS $\cA$ that presents some monoid $M$, we say that $\cA$ is split to mean that, for any choice of isomorphism $\theta:Q(A) \to M$, the presentation $(\cA,\theta)$ is split.
\end{remark}

Next, we show that being slip is forward invariant under performing GETTs.

\begin{proposition}\label{proposition: Split is forward property}
    Let $\cA$ and $\cB$ be two presentations of a monoid $M$ such that $\cA \rightsquigarrow \cB$. If $\cA$ is split, then so is $\cB$.
\end{proposition}
\begin{proof}
    Consider the following commutative
    \begin{center}
        \begin{tikzcd}[column sep=5]
            A\arrow[rr, "f"]\arrow[d, "q_A", swap]& & B\arrow[d, "q_B"] \\
            Q(A)\arrow[rr, "\theta"]\arrow[dr, "\alpha", swap]& & Q(B) \arrow[dl, "\beta"] \\
            & M \arrow[luu, "s_\alpha", bend left=90] \arrow[ruu, dashed, bend right=90, "?", swap]&
        \end{tikzcd}
    \end{center}

    Where the commutative square is given by Theorem \ref{theorem: if square commutes then GETT sequence}, and $\alpha$ and $\beta$ are isomorphisms chosen so that the triangle commutes. Define $s_\beta:= fs_\alpha \alpha \theta^{-1}\beta^{-1}$. Then
    \begin{align*}
        \beta q_Bs_\beta &= \beta (q_B f)s_\alpha \alpha \theta^{-1}\beta^{-1} \\
        &= (\beta \theta) q_As_\alpha \alpha \theta^{-1}\beta^{-1} \\
        &= (\alpha q_As_\alpha) \alpha \theta^{-1}\beta^{-1} \\
        &= \alpha \theta^{-1}\beta^{-1} \\
        &= 1_M
    \end{align*}

    Which concludes the proof.
\end{proof}

Note that the converse of this proposition is not true, in the sense that we may start with a non-split MRS $\cA$ and by performing GETTs, we may obtain a MRS $\cB$ which is split, as per the next example.

\begin{example}\normalfont
Let $M=(\Z_2,+)$ and consider the following two presentations
\[
    G(\Z_2)=(F_{\{1\}},\oplus,R)
    \qquad\text{and}\qquad
    D(\Z_2)=(\Z_2,+,\emptyset).
\]
Where $R = \{(11,\varepsilon), (1,1)\}$. Under the canonical identification $Q(G(\Z_2))\simeq\Z_2$, the quotient map $q:F_{\{1\}}\twoheadrightarrow\Z_2$ sends $1^n$ to $n \pmod 2$.

We start by showing that $\cA$ is not split via a contradiction. Assume that $q$ admits a section $s:\Z_2\to F_{\{1\}}$. Since $s$ is a monoid homomorphism,
\[
    s(1)\oplus s(1)=s(1+1)=s(0)=\varepsilon.
\]
From this, we conclude that $s(1)$ must be the empty word, since otherwise it would be impossible for $s(1)\oplus s(1)$ to be empty. This contradicts the hypothesis that $q(s(1))=1$. Hence $G(M)$ is not split.

On the other hand, $D(M)$ is trivially split and we know that $G(M) \rightsquigarrow D(M)$ as $G(M)$ is weakly initial.
\end{example}

Being split is purely an algebraic property of an MRS. This property, however, also manifests itself in terms of rewriting properties as given in the following proposition.

\begin{proposition}\label{Proposition: Split iff can get from D(M)}
    Let $\cA$ be a presentation of $M$. Then $\cA$ splits if and only if $D(M) \rightsquigarrow \cA$.
\end{proposition}
\begin{proof}
    Because $D(M)$ is trivially split, if $D(M) \rightsquigarrow \cA$, then it follows by Proposition \ref{proposition: Split is forward property} that $\cA$ is also split.

    On the other hand, assume that $\cA$ is split. Let $q_A : A \to Q(A)$ denote the projection, and fix $\theta: Q(A) \to M$ to be any isomorphism. Then the following diagram commutes

    \begin{center}
        \begin{tikzcd}
             D(M) \arrow[r, "s_\theta"]\arrow[d, "\mathrm{id}", swap] & A \arrow[d, "q_A"]\\ M\arrow[r, "\theta^{-1}", swap] & Q(A)
        \end{tikzcd}
    \end{center}
    by definition, as $s_\theta$ satisfies the property $\theta q_As_\theta = 1_M$. From Theorem \ref{theorem: if square commutes then GETT sequence} if follows that $D(M) \rightsquigarrow \cA$.
\end{proof}

The following theorem says that we can detect then a monoid $M$ is free purely by the rewriting theoretic behavior of its presentations.

\begin{theorem}\label{theorem caracterization of free monoids}
Let $M$ be a monoid. The following are equivalent:

\begin{enumerate}
    \item $M$ is a free monoid;
    \item Every presentation of $M$ is split;
    \item $D(M)\rightsquigarrow G(M)$
    \item Every presentation of $M$ can be obtained from any others via a sequence of GETTs.
\end{enumerate}
\end{theorem}
\begin{proof}
    $(1) \Rightarrow (2)$ Let $M$ be a free monoid, say $M = F_X$ for some set $X$, and let $\cA$ be a presentation of $M$. Fix an isomorphism $\theta: Q(A) \to M$, let $q_A:A \to Q(A)$ denote the canonical projection and $q_A^\theta: A \rightarrow M$ denote $\theta q_A$. For each $x \in X$, let $a_x$ be an element in $A$ such that $ q_A^\theta(a_x) = x$, which exists by the surjectivity of $q_A^\theta$. As $M$ is the free monoid generated by $X$, the assignment $x \mapsto a_x$ extends to a monoid homomorphism $s:M \to A$. By construction, this monoid homomorphism satisfies $q_A^\theta s = 1_M$.

    $(2) \Rightarrow(3)$ If every presentation of $M$ is split, then in particular $G(M)$ is split. By Proposition \ref{Proposition: Split iff can get from D(M)}, we conclude that $D(M) \rightsquigarrow G(M)$.

    $(3) \Rightarrow (4)$ Let $\cA$ and $\cB$ be two presentations of $M$. as $D(M)$ is weakly terminal and $G(M)$ weakly initial in $\pres(M)$, we conclude that $\cA \rightsquigarrow D(M)$ and $G(M) \rightsquigarrow \cB$. The fact that $\cA \rightsquigarrow \cB$ follows then by our hypothesis that $D(M) \rightsquigarrow G(M)$.

    $(4) \Rightarrow (1)$ For a monoid $M$, hypothesis $(4)$ implies, in particular, that $D(M) \rightsquigarrow G(M)$. By Proposition \ref{Proposition: Split iff can get from D(M)}, this means that $G(M)$ is split, i.e. there exists a monoid homomorphism $s:M \to F_{M^+}$ such that $\mathrm{e}_M s = 1_M$, where $\mathrm{e}_M: F_{M^+} \to M$ is the projection map. In other words, $M$ is a retract of $F_{M^+}$. By Corollary 3 in \cite{noauthor_expanded_nodate}, the retract of any free monoid is also free, meaning that $M$ is free.
\end{proof}

We finish this section off with a summary of its results. The motivation was to provide an answer to the following question:  Given a fixed monoid $M$, what can be said about the MRSs that present it? What is the relationship between two presentations of the same monoid? 

In general, given an MRS $\cA$ that presents $M$, we have shown that $$G(M) \rightsquigarrow \cA \rightsquigarrow D(M),$$ and this property gives a precise characterization of every MRS that presents $M$.

As per the relationship between two MRS $\cA$ and $\cB$ that present the same monoid $M$, in general the best relationship we can derive is that we have the following square in $\pres(M)$:
\begin{center}
    \begin{tikzcd}[row sep=10, column sep=10]
        & G(M)\arrow[dl,rightsquigarrow ]\arrow[dr,rightsquigarrow] & \\
        \cA\arrow[dr,rightsquigarrow] & &\cB\arrow[dl,rightsquigarrow] \\
        & D(M)&
    \end{tikzcd}
\end{center}

The following example constructs a monoid $M$ and two presentations $\cA$ and $\cB$ such that $\cA \not \rightsquigarrow \cB$ and $\cB \not \rightsquigarrow \cA$, to show that in general, we cannot move from one presentation to the other using GETTs.

\begin{example}\normalfont
For $k\in\{1,2\}$, let
\[
    T_k=(\{0,1,\ldots,k\},\oplus_k),
    \qquad
    i\oplus_k j=\min\{i+j,k\},
\]
and consider the monoid $M=T_1\times T_2$. Define
\[
    A=T_1\times \N,
    \qquad
    R=\left\{\bigl((0,3),(0,2)\bigr)\right\},
\]
and
\[
    B=\N\times T_2,
    \qquad
    L=\left\{\bigl((2,0),(1,0)\bigr)\right\}.
\]
Let $\cA=(A,R)$ and $\cB=(B,L)$, and note that both are presentations for the monoid $M$.

The corresponding quotient maps are
\[
    q_A:A\longrightarrow M,
    \qquad
    q_A(i,n)=\bigl(i,\min\{n,2\}\bigr),
\]
and
\[
    q_B:B\longrightarrow M,
    \qquad
    q_B(n,j)=\bigl(\min\{n,1\},j\bigr).
\]
Indeed, the unique $R$-irreducible form of $(i,n)\in A$ is $\bigl(i,\min\{n,2\}\bigr)$, while the unique $L$-irreducible form of $(n,j)\in B$ is $\bigl(\min\{n,1\},j\bigr).$
In particular, both $\cA$ and $\cB$ are Noetherian and confluent.

We claim that $\cA \not \rightsquigarrow \cB$ and $\cB \not \rightsquigarrow \cA$. First observe that the idempotents of $M$ are
\[
    (0,0),\qquad (1,0),\qquad (0,2),\qquad (1,2).
\]
Every automorphism of $M$ fixes $(0,0)$, since it is the identity, and
$(1,2)$, since it is the unique absorbing element. Moreover, $(0,2)$
admits a non-idempotent square root, namely $(0,1)$:
\[
    (0,1)\oplus(0,1)=(0,2),
\]
whereas $(1,0)$ does not. As such, it follows that every automorphism of $M$ fixes
both $(1,0)$ and $(0,2)$.

Suppose that $\cA\rightsquigarrow\cB$. By Theorem
\ref{theorem: if square commutes then GETT sequence}, there would exist a
monoid homomorphism $f:A\to B$ and an automorphism $\alpha:M\to M$ such
that $q_Bf=\alpha q_A.$

The element $(1,0)\in A$ is idempotent and satisfies
$q_A(1,0)=(1,0)$. Therefore $f(1,0)$ would have to be an idempotent
belonging to
\[
    q_B^{-1}(1,0)=\{(n,0):n\geq 1\}.
\]
However, none of these elements is idempotent, since
\[
    (n,0)+(n,0)=(2n,0)\neq(n,0)
\]
for every $n\geq1$. Hence $\cA\not\rightsquigarrow\cB$. An analogous argument shows that $\cB \not \rightsquigarrow \cA$.
\end{example}

By Theorem \ref{theorem caracterization of free monoids}, the property that every presentation can be obtained from any other using GETTs fully characterizes the fact that the monoid being presented is free.

However, we have shown that every presentation as a string rewriting system is weakly initial in $\pres(M)$. From this fact, we can recover the following generalization of the classical Tietze Theorem:

\begin{corollary}
    Let $M$ be a monoid and $\cA$,$\cB$ be two string rewriting system that present $M$. Then $\cA \rightsquigarrow \cB$ and $\cB \rightsquigarrow \cA$.
\end{corollary}

Regarding the category $\pres(M)$, we have also proved in Proposition \ref{Proposition: Split iff can get from D(M)} a characterization of the weakly terminal objects:

\begin{corollary}
    The weakly terminal objects in $\pres(M)$ are precisely the split presentations of $M$.
\end{corollary}

Regarding the weakly initial objects, their behavior is not so clear: given a monoid $M$, Proposition \ref{proposition: SRS are initial} tells us that every presentation as a string rewriting system is weakly initial in $\pres(M)$. However, the converse is not true: there may be presentations that are weakly initial but whose base monoid is not free, as per the following example.

\begin{example}\normalfont
Let $E=\{1_E,e\}$ be a monoid with two elements such that $e^2=e$, and consider the monoid $A=F_{\{1\}}\times E$. Let $\cA=(A,R)$, where
\[
R=
\left\{
\bigl((1\oplus1,1_E),(\varepsilon,1_E)\bigr),
\bigl((\varepsilon,e),(\varepsilon,1_E)\bigr)
\right\}.
\]
Every element of $A$ is of the form $(1^n,\delta)$, with $n\in\N$ and
$\delta\in E$, and it rewrites to
\[
    (1^n,\delta)\to_R^*
    \bigl(1^{\,n\bmod 2},1_E\bigr),
\]
where $1^0=\varepsilon$. These are the unique irreducible forms and as such, its easy to see that $\cA$ presents $\Z_2$. Under the canonical
isomorphism $Q(\cA)\cong \Z_2$, its quotient map is given by
\[
    q_A(1^n,\delta)=n\pmod 2.
\]

Note also that $A$ is not free as it has an idempotent: $(\varepsilon,e)$. We claim that $\cA$ is, never the less, weakly initial in $\pres(\Z_2)$.
To see this, consider the projection
\[
    p:F_{\{1\}}\times E\longrightarrow F_{\{1\}}
\]
onto the first coordinate. This is a monoid homomorphism and, in particular, the following diagram commutes:
\[
\begin{tikzcd}
F_{\{1\}}\times E
    \arrow[r,"p"]
    \arrow[d,"q_A",swap]
&
F_{\{1\}}
    \arrow[d,"q_{G(\Z_2)}"]
\\
\Z_2
    \arrow[r,"\mathrm{id}",swap]
&
\Z_2.
\end{tikzcd}
\]
By Theorem \ref{theorem: if square commutes then GETT sequence}, it follows
that $\cA\rightsquigarrow G(\Z_2)$. Since $G(\Z_2)$ is weakly initial
in $\pres(\Z_2)$, then so is $\cA$, even though $A$ is not free.
\end{example}

As such, the following questions still remains:
\begin{question}\normalfont
    Given a monoid $M$, what property characterizes the weakly initial objects in $\pres(M)$?
\end{question}
 At present, the author is unable to provide and answer to this question.

\section*{Appendix A}

The aim of this appendix is to provide a comprehensive proof of Theorem \ref{biadju}.

\begin{theorem*}
    The pair $(G,I)$ forms a 2-adjunction, where $G$ is left adjoint and $I$ is right adjoint.
    \[\begin{tikzcd}
\cat{Mon}\ar[r,bend left,"G",""{name=A, below}] & \cat{NCRS_2} \ar[l,bend left,"I",""{name=B,above}] \ar[from=A, to=B, "\bot", phantom]
\end{tikzcd}\]

In particular, the unit $\eta:1_{\cat{Mon}} \to IG$ is the identity 2-natural transformation.
\end{theorem*}

\begin{proof}
    Start by noting that $IG = 1_\cat{Mon}$ as (strict) 2-functors. Indeed, given a monoid $(M,\cdot)$, then the monoid of irreducibles $(\overline{F_{M^+}}, \bar \oplus)$ is equal to $(M,\cdot)$.

    Additionally, let $\phi:(M,\cdot) \to (N,*)$ be a monoid homomorphism and $u \in M$.
    Then
\[
IG\phi(u)=I(G\phi)(u)=\overline{G\phi(u)}=\overline{\nu_N(\phi(u))}=\nu_N(\phi(u)),
\]
and under the identification $\nu_N(N)\subseteq F_{N^+}$ this is just $\phi(u)\in N$.
   
    As such, with respect to the hom-categories of $\cat{Mon}$, $$IG: \hom_{\cat{Mon}}(M,N) \to \hom_{\cat{Mon}}(IG(M), IG(N))$$ is the identity functor.

    Because of this, we consider the unit $\eta:  1_{\cat{Mon}} \to IG $ to be the identity strict 2-natural transformation.

    On the other hand, we define the counit $$\varepsilon:GI \to  1_{\cat{NCRS_2}}$$
    to be the strong transformation that, to each object $(A,\cdot , R)$ associates the $1$-cell:
    \begin{align*}
        \varepsilon_A: GI(A,\cdot, R) &\to (A,\cdot, R)\\
        a_1 \oplus \ldots \oplus a_n & \mapsto a_1\cdot\ldots\cdot a_n
    \end{align*}
    where $a_i \in (\overline{A})^+$ are letters in $F_{(\overline{A})^+}$, and  $\varepsilon_A(\varepsilon) = 1_A$, where $\varepsilon$ denotes the empty word.

    Additionally, for each 1-cell $\phi:(A,\cdot, R) \to (B,*,L)$ , we associate the unique 2-cell

        \begin{center}
\begin{tikzcd}
    GI(A,\cdot, R) 
        \arrow[r, "\varepsilon_A"] 
        \arrow[d, "GI\phi"'] 
    & (A,\cdot, R) 
        \arrow[d, "\phi"] 
        \arrow[dl, Rightarrow, shorten <=3pt, shorten >=3pt] \\
    GI(B, *,L)
        \arrow[r, "\varepsilon_B"'] 
    & (B, *,L)
\end{tikzcd}
\end{center}

Indeed, given a string $a_1 \oplus \ldots \oplus a_n \in GI(A,\cdot, R)$ with each $a_i \in (\overline{A})^+$, we have that $$(\phi \circ \varepsilon_A)(a_1 \oplus \ldots \oplus a_n) = \phi(a_1\cdot \ldots \cdot a_n)=\phi(a_1)* \ldots* \phi(a_n)$$
$$(\varepsilon_B\circ GI\phi)(a_1 \oplus \ldots \oplus a_n) = \overline{\phi(a_1)}* \ldots* \overline{\phi(a_n)}$$

Now note that \begin{align*}
    \overline{\phi(a_1)*  \ldots* \phi(a_n)}
    &= \overline{\overline{\phi(a_1)}* \ldots* \overline{\phi(a_n)}}
\end{align*}
So $$(\phi \circ \varepsilon_A)(a_1 \oplus \ldots \oplus a_n) \leftrightarrow_L^* (\varepsilon_B\circ GI\phi)(a_1 \oplus \ldots \oplus a_n)$$
meaning indeed there exists a unique 2-cell $\phi \circ \varepsilon_A \Rightarrow \varepsilon_B\circ GI\phi$.

We start by verifying that $\varepsilon$ is well defined.

\vspace{10pt}

    \textbf{Claim:} For all Noetherian confluent MRS, the map $\varepsilon_A: GI(A,\cdot, R) \to (A,\cdot, R)$ is a MRS-homomorphism and therefore a 1-cell in $\cat{NCRS_2}$.

    We start by first showing that $\varepsilon_A:(F_{(\overline{A})^+}, \oplus) \to (A,\cdot)$ is a monoid homomorphism. Indeed, $\varepsilon_A$ equals the composition of the homomorphisms $\ev \circ \iota$:
    \begin{center}
        \begin{tikzcd}
            (F_{(\overline{A})^+}, \oplus) \arrow[r, hook, "\iota"]& (F_{
            A^+}, \oplus) \arrow[r, "\ev", dashed]& (A, \cdot)\\
        \end{tikzcd}
    \end{center}
    Where $\ev$ is defined via the universal property of free monoids when applied to the inclusion $ A^+ \hookrightarrow A$.

    Additionally, to show that $\varepsilon_A$ is compatible with rewriting, let $(a\oplus b,c) \in R_{\overline{A}}$ for some $a,b,c \in \overline A$. 
    
    We wish to show that $ab \to_R^* c$ in $(A,\cdot,R)$. By definition, the fact that $(a\oplus b,c) \in R_{\overline{A}}$ tells us that $a \ \bar \cdot \ b = c$ in $(\overline A, \bar \cdot)$, which means that $\overline{a b } = c$ from where it follows that $ab \to_R^* c$. Hence, $\varepsilon_A$ is a MRS-homomorphism.

    \vspace{10pt}
    
    \textbf{Claim:} $\varepsilon$ is a strong transformation.

    Consider a $2$-cell 
    
    \begin{center}
        \begin{tikzcd}
            (A,\cdot, R) \arrow[r, "\phi", bend left, ""{name=lF, below}]  \arrow[r, "\psi", bend right, swap, ""{name=lHup}]& (B,*,L) \arrow[Rightarrow, from=lF, to=lHup,  end anchor={[yshift=3pt]}] 
        \end{tikzcd}
    \end{center}

    Then the following equality of pasting diagrams hold trivially, because the hom-categories in $\cat{NCRS_2}$ are thin.
    \begin{center}
        \begin{tikzcd}
            GI(A,\cdot, R) \arrow[dd, bend left=50, "GI \phi"description, ""{name=U}] \arrow[dd,bend right=50, "GI \psi"description, ""{name=D} ]\arrow[Rightarrow, from=U,to=D, end anchor={[xshift=5pt]}, start anchor={[xshift=-10pt]}] \arrow[rr, "\varepsilon_A" description] && (A,\cdot, R)\arrow[dd, bend left =50, "\phi"description] \arrow[Rightarrow, bend left=10, ddll]\\
            &&\\
            GI(B,*, L)\arrow[rr, "\varepsilon_B" description]&&  (B,*, L)
        \end{tikzcd}\hspace{10pt} = \hspace{5pt}
        \begin{tikzcd}
            GI(A,\cdot, R)\arrow[rr, "\varepsilon_A" description] \arrow[dd, bend right=50pt, "GI\psi"description] && (A,\cdot, R)\arrow[ddll, bend right=10, Rightarrow] \arrow[dd, bend left=50, "\phi"description, ""{name=U}] \arrow[dd,bend right=50, "\psi"description, ""{name=D} ]\arrow[Rightarrow, from=U,to=D, end anchor={[xshift=3pt]}, start anchor={[xshift=-5pt]}] \\
            && \\
            GI(B,*, L) \arrow[rr, "\varepsilon_B" description]&&  (B,*, L)
        \end{tikzcd}
    \end{center}

\vspace{10pt}

Finally, to conclude the proof we need to prove the following: 

 \textbf{Claim:} $\eta$ and $\varepsilon$ satisfy the triangle identities.

        Start by consider the following diagram:

        \begin{center}
            \begin{tikzcd}
                I \arrow[dr, "1_I", swap]\arrow[r, "\eta I"] &IGI \arrow[d, "I\varepsilon"]\\
                & I
            \end{tikzcd}
        \end{center}
         To see that $I\varepsilon\circ \eta I = 1_I$, we start by verifying that such is the case for the components indexed by objects.
        Let $(A,\cdot, R)$ be a Noetherian confluent MRS, and consider the diagram: 
        \begin{center}
            \begin{tikzcd}
                I(A,\cdot, R) \arrow[dr, "1_{I (A)}", swap]\arrow[r, "\eta_{I(A)}"] &IGI(A,\cdot, R) \arrow[d, "I(\varepsilon_A)"]\\
                & I(A,\cdot, R)
            \end{tikzcd}
        \end{center}

        Let $a \in \overline A$ be any irreducible element and note that $(I(\varepsilon_A)\circ \eta_{I(A)})(a) = a$. As $a \in \overline{A}$ is arbitrary, we get that $I(\varepsilon_A)\circ \eta_{I(A)} = (1_A)_a$. On the other hand, consider a MRS-homomorphism $\phi:(A,\cdot, R) \to (B,*,L)$. Then:

        \begin{center}

                \begin{tikzcd}
                I(A,\cdot, R)\arrow[r, "\eta_{I(A)}"]\arrow[dd, "I\phi"description] & IGI(A,\cdot, R)\arrow[r, "I(\varepsilon_A)"] \arrow[Rightarrow, ddl, "\eta I(\phi)"description] \arrow[dd, "IGI\phi"description] & I(A,\cdot, R) \arrow[dd, "I\phi"description]\arrow[Rightarrow, ddl, "I\varepsilon(\phi)"description]\\
                &&\\
                I(B,*,L)\arrow[r, "\eta_{I(B)}", swap] &IGI(B,*,L) \arrow[r, "I(\varepsilon_B)", swap] & I(B,*,L)\\
            \end{tikzcd}

            =
            
\vspace{10pt}
            \begin{tikzcd}
                I(A,\cdot, R)\arrow[rr, "1_{I(A)}"]\arrow[dd, "I\phi"description] &  & I(A,\cdot, R) \arrow[dd, "I\phi"description]\arrow[Rightarrow, ddll, "1_I(\phi)"description]\\
                &&\\
                I(B,*,L)\arrow[rr, "1_{I(B)}", swap] & & I(B,*,L)\\
            \end{tikzcd}
        \end{center}

        Meaning that $I\varepsilon\circ \eta I$ agrees with $1_I$ on the components index by 1-cells.

        On the other hand, consider the triangle:
        \begin{center}
            \begin{tikzcd}
                G \arrow[dr, "1_G", swap]\arrow[r, "G\eta"] &GIG \arrow[d, "\varepsilon G"]\\
                & G
            \end{tikzcd}
        \end{center}

        Note that given a monoid $(M,\cdot)$ and a word $m_1 \oplus \ldots \oplus m_n$ in $G(M)$, then $$(\varepsilon_{G(M)} \circ G(\eta_M))(m_1 \oplus \ldots \oplus m_n) = m_1 \oplus \ldots \oplus m_n$$ meaning that $\varepsilon G \circ G\eta$ coincides with $1_G$ on the components indexed by objects.

        A similar argument as for the previous triangle shows that $\varepsilon G \circ G\eta$ also coincides with $1_G$ on the components indexed by $1$-cells, i.e. the triangle commutes strictly. 
\end{proof}

\section{Acknowledgments}
The author would like to thank Pedro V. Silva for his feedback and support during the elaboration of this work.

\bibliographystyle{unsrt}
\bibliography{references}

@article{power_thues_2013,
  title={Thue's 1914 paper: a translation},
  author={Power, James F},
  journal={arXiv preprint arXiv:1308.5858},
  year={2013}
}

@boook{book_string-rewriting_1993,
  title={String-rewriting systems},
  author={Book, Ronald V and Otto, Friedrich},
  booktitle={String-Rewriting Systems},
  year={1993},
  publisher={Springer}
}

@book{terese_term_2003,
  title={Term rewriting systems},
  author={Bezem, Marc and Klop, Jan Willem and de Vrijer, Roel},
  year={2003},
  publisher={Cambridge University Press}
}

@book{johnson_2-dimensional_2021,
    address = {Oxford, New York},
    title = {2-{Dimensional} {Categories}},
    isbn = {978-0-19-887138-5},
    publisher = {Oxford University Press},
    author = {Johnson, Niles and Yau, Donald},
    month = jan,
    year = {2021},
}

@article{burroni_higher-dimensional_1993,
    title = {Higher-dimensional word problems with applications to equational logic},
    volume = {115},
    issn = {0304-3975},
    url = {https://doi.org/10.1016/0304-3975(93)90054-W},
    doi = {10.1016/0304-3975(93)90054-W},
    number = {1},
    urldate = {2025-12-09},
    journal = {Theor. Comput. Sci.},
    author = {Burroni, Albert},
    month = jul,
    year = {1993},
    pages = {43--62},
}

@book{caramello_theories_2017,
    title = {Theories, {Sites}, {Toposes}: {Relating} and studying mathematical theories through topos-theoretic 'bridges'},
    isbn = {978-0-19-875891-4},
    shorttitle = {Theories, {Sites}, {Toposes}},
    url = {https://doi.org/10.1093/oso/9780198758914.001.0001},
    urldate = {2025-12-11},
    publisher = {Oxford University Press},
    author = {Caramello, Olivia},
    month = dec,
    year = {2017},
    doi = {10.1093/oso/9780198758914.001.0001},
}

@incollection{lack_2-categories_2010,
    address = {New York, NY},
    title = {A 2-{Categories} {Companion}},
    isbn = {978-1-4419-1524-5},
    url = {https://doi.org/10.1007/978-1-4419-1524-5_4},
    language = {en},
    urldate = {2025-12-16},
    booktitle = {Towards {Higher} {Categories}},
    publisher = {Springer},
    author = {Lack, Stephen},
    editor = {Baez, John C. and May, J. Peter},
    year = {2010},
    doi = {10.1007/978-1-4419-1524-5_4},
    pages = {105--191},
}

@book{baader_term_1998,
    address = {Cambridge},
    title = {Term {Rewriting} and {All} {That}},
    isbn = {978-0-521-77920-3},
    url = {https://www.cambridge.org/core/books/term-rewriting-and-all-that/71768055278D0DEF4FFC74722DE0D707},
    urldate = {2026-01-12},
    publisher = {Cambridge University Press},
    author = {Baader, Franz and Nipkow, Tobias},
    year = {1998},
    doi = {10.1017/CBO9781139172752},
}

@book{ara_polygraphs_2025,
    address = {Cambridge},
    series = {London {Mathematical} {Society} {Lecture} {Note} {Series}},
    title = {Polygraphs: {From} {Rewriting} to {Higher} {Categories}},
    isbn = {978-1-009-49898-2},
    shorttitle = {Polygraphs},
    url = {https://www.cambridge.org/core/books/polygraphs-from-rewriting-to-higher-categories/5B8F3783067A2303C3A945ED6CA7F4C0},
    urldate = {2026-01-15},
    publisher = {Cambridge University Press},
    author = {Ara, Dimitri and Burroni, Albert and Guiraud, Yves and Malbos, Philippe and Métayer, François and Mimram, Samuel},
    year = {2025},
    doi = {10.1017/9781009498968},
}

@book{noauthor_course_nodate,
  author    = {Burris, Stanley and Sankappanavar, H. P.},
  title     = {A Course in Universal Algebra},
  series    = {Graduate Texts in Mathematics},
  volume    = {78},
  publisher = {Springer-Verlag},
  address   = {New York},
  year      = {1981}
}

@article{noauthor_expanded_nodate,
  author  = {Head, Tom},
  title   = {Expanded subalphabets in the theories of languages and semigroups},
  journal = {International Journal of Computer Mathematics},
  volume  = {12},
  number  = {2},
  pages   = {113--123},
  year    = {1982},
  doi     = {10.1080/00207168208803330}
}

\end{document}